\documentclass{article}
\usepackage[numbers,sort&compress]{natbib}
\usepackage{amsbsy} %\boldsymbol command for vectors; see AMS conventions
\usepackage{booktabs}	%for neater, profession tables
	\usepackage{multirow} %not yet tested 
	\usepackage{multicol}
\usepackage{comment}  %Required by Elsarticle. Load explicitly to make available for drafts
\usepackage{color}
\usepackage{multicol}
\usepackage{graphicx}
\graphicspath{{./}} %upload
\usepackage[version=3]{mhchem} % Chemical symbols
\usepackage{setspace}
\usepackage{subfigure}
\usepackage{url}
\usepackage{xspace}
\usepackage{lastpage} % for page count 
\usepackage[left=2.5cm,bottom=2.5cm,top=2.5cm,right=2.5cm]{geometry}

\usepackage{fancyhdr}
\fancypagestyle{plain}{ %
\fancyhf{} % remove everything
\setlength{\headheight}{15.2pt}
\setlength{\footskip}{40pt}
\lhead{\textit{The effect of precipitation on a Mg-Zn-Y alloy}}
\rhead{J.M. Rosalie, H. Somekawa, A. Singh, T. Mukai}
\lfoot{\textit{Accepted author manuscript}}
\rfoot{\thepage\ of\ \pageref{LastPage}}

}

\usepackage[plainpages=false,
pdftitle={The effect of precipitation on strength and ductility in a Mg-Zn-Y alloy},
pdfauthor={J.M. Rosalie,H. Somekawa, A. Singh, T. Mukai},
pdfsubject={Metals and alloys; Precipitation; Microstructure; Transmission electron microscopy},
pdfkeywords={Aluminium alloys, Precipitation, Interfaces,Segregation},
colorlinks,
debug,
linkcolor=red,
citecolor=red, 
filecolor=blue,
breaklinks=true]{hyperref}
\newcommand{\betap}{\ensuremath{{\beta_1^\prime}}\xspace}
\newcommand{\micron}{\ensuremath{\mathrm{\mu m}}\xspace}
\newcommand{\celsius}{\ensuremath{\mathrm{^oC}}\xspace} %degrees Celsius

\newcommand{\fig}[2][6cm]{\resizebox{#1}{!}{\includegraphics{#2}}} %variable size figure
\begin{document}

%********************************************* FRONT MATTER ********************%

\title{The effect of precipitation on strength and ductility in a Mg-Zn-Y alloy} 
\date{}

\author{Julian~M~Rosalie$^{a}$ \and 
Hidetoshi Somekawa $^{a}$\and 
Alok Singh $^{a}$ \and Toshiji Mukai $^{b}$}

\maketitle
\noindent
$^a$Structural Materials Unit, National Institute for Materials Science, Tsukuba, 305-0047, Japan.\\
$^b$Dept. Mechanical Engineering, Kobe University, 1-1 Rokkodai, Nada, Kobe city, 657-8501 Japan.

%************************* 	ABSTRACT 		****************************************%

\begin{abstract} 

The effect of pre-ageing deformation on the size and distribution of \betap precipitates and subsequently on the resulting strength and ductility have been measured in a Mg-3.0at.\%Zn-0.5at.\%Y alloy.
The alloy was extruded and then subjected to a T8 heat treatment comprised of a solution-treatment, cold-work and artificial ageing.
Extrusion was used to introduce texture, ensuring that deformation occurred via slip rather than twinning. 
Samples were subjected to controlled uniaxial deformation and then isothermally aged to peak hardness. 
Precipitate length, diameter and number density were measured and evaluated in terms of the strength and ductility of the alloy. 
The nucleation of the \betap precipitates in peak-aged condition without pre-ageing deformation (i.e.T6 treatment) was poor, with only 0.5\% volume fraction, compared to  approximately 3.5\% in T6 treated binary Mg-3.0at.\%Zn alloy. 
The microstructure of the Mg-Zn-Y alloy was less refined, with  larger diameter  precipitates and lower \betap number densities compared to a binary Mg-3.0at.\%Zn alloy. 
Deformation to 5\% plastic strain increased the volume fraction of \betap precipitates to approximately 2.3\% and refined the \betap precipitate length and diameter. 
The combination of these effects increased the yield strength after isothermal ageing from 217\,MPa (0\% cold-work) to 287\,MPa (5\% cold-work).  
The yield stress increased linearly with reciprocal interparticle spacing on the basal and prismatic planes and the alloy showed similar strengthening against basal slip to Mg-Zn.  
The elongation  increased linearly with particle spacing. 
The ductility of Mg-Zn-Y alloys was similar to that of Mg-Zn for equivalently spaced particles.

\end{abstract}
%*************************************************************************************%

%********************************* Keywords ********************************************%
\paragraph{Keywords} 
Metals and alloys; Precipitation; Microstructure; Transmission electron microscopy

%elsarticle
%************************************end of front matter ************************************%

\section{Introduction}

Magnesium alloys are valued in transport applications, where their low density offers substantial weight savings compared to steel or aluminium. 
The addition of yttrium and rare earth elements (RE) to magnesium alloys is widespread in alloys where resistance to creep or  corrosion is required, such as in the WE series of alloys \cite{KielbusElektron2007, WangCreep2001, LuoReview2004,BambergerReview2008}.

The use of RE and Yttrium additions has been extended to a range of other Mg alloy systems.
These systems include alloys where the principle alloying element is zinc, such as the ZK (magnesium-zinc-zirconium) series.
In addition to improved creep properties \cite{Boehlert2007}, the addition of rare earth elements weakens the texture development during extrusion and rolling of Mg alloys, reducing the  tensile-compressive asymmetry \cite{BallPrangnell1994,BohlenTexture2007,StanfordTexture2008}.
The presence of high-melting grain boundary phases can also limit coarsening \cite{Li-Y-2011}. 

Wrought Mg alloys containing zinc are strengthened by a binary precipitate, termed \betap, that forms extended rods parallel to the hexagonal axis.  
In a recent study on precipitation hardening of Mg-Zn alloys,  we  found that the yield strength of 273\,MPa 
for T6 heat treated material (solution treatment plus artificial ageing) could be increased to 309\,MPa by using a T8 treatment (solution treatment, cold-work and artificial ageing). 
This was due to dislocations providing heterogeneous nucleation sites for \betap precipitates, leading to accelerated kinetics and a refinement of the precipitate size \cite{RosalieMgZn2012}. 

While it has been reported that RE additions reduced the rate of overageing in Mg-Zn alloys \cite{WeiPrec1995}, there have been no systematic studies on precipitation strengthening of Mg-Zn-Y alloys. 
Yttrium has virtually zero solubility in magnesium at room temperature \cite{ShaoCalphad2006} 
 and partitions to the grain boundaries in binary Mg-Y alloys \cite{HadornTexture2012} and to
ternary phase precipitates in Mg-Zn-Y alloys \cite{WeiPrec1995,Singh2007,Tsai1994}. 
The intragranular region remains low in Y and Mg-Zn-Y alloys therefore show a similar precipitation sequence to binary Mg-Zn.
For  Zn:Y ratios of approximately 6:1 \cite{TsaiPhaseDiagram2000}, Mg-Zn-Y alloys precipitate a quasicrystalline phase (i-phase, \ce{Mg3Zn6Y}) \cite{Tsai1994}. 
This phase exists in two-phase equilibrium with the matrix at high temperatures and precipitates as a grain boundary eutectic during solidification  \cite{TsaiPhaseDiagram2000}.  
In alloys with Zn:Y ratios close to the stoichiometry of the i-phase, therefore, little zinc is expected to be available for \betap phase precipitation and the precipitation hardening response is expected to be poor.

This study set out to determine whether cold work prior to ageing  could stimulate \betap precipitation in a wrought Mg-Zn-Y alloy and to quantify the strength and ductility as a function of the precipitate size and distribution.
Deformation twinning was prevented by extruding the alloy to develop textures favourable for slip rather than  twinning.
Controlled deformation was applied prior to isothermal ageing to stimulate \betap precipitate nucleation.
Varying the amount of pre-ageing deformation altered the size and distribution of the \betap precipitates and thus affected the yield strength and ductility in the peak-aged condition. 
 
\section{Experimental details}

Billets of Mg-3.0at.\%Zn-0.5at.\%Y alloys were prepared from pure elements via direct-chill casting. 
The compositions were confirmed via inductively-coupled plasma mass spectroscopy (ICP-MS). 
The billets were homogenised for 15\,h at 350$^\circ$C and extruded into 12\,mm diameter rods at 300$^\circ$C with an extrusion ratio of 12:1.
Tensile samples of gauge length 15\,mm and diameter 3\,mm were machined from the extruded rods.  

The tensile samples were  given a T8 heat treatment comprised of a solution-treatment followed by varying levels of cold-work and finally artificial ageing. 
Samples were encapsulated in helium, solution treated for 1\,h at 400$^\circ$C and and quenched into water. 
Pre-ageing deformation was carried out using an Instron mechanical tester, to a nominal plastic strain of 3\% or 5\%. 
Tensile deformation was applied parallel to the extrusion axis at a strain rate, $\dot{\epsilon}$, of \(1\!\times\!10^{-3}\,\mbox{s}^{-1}\).  
Sections were polished and etched with acetic picral for examination by light microscopy. 
The remaining samples were then aged to peak hardness in an oil bath at 150\celsius for periods of 216-256\,h for non-deformed samples and 32-48\,h for deformed samples.
The ageing response was measured using Vickers hardness testing with a 300\,g load. 
Tensile tests were carried out in the Instron mechanical tester ($\dot{\epsilon}=1\!\times\!10^{-3}\,\mbox{s}^{-1})$ on samples aged to peak hardness. 
Solution-treated and quenched samples (10\,mm height, 5\,mm diameter)  were also prepared and tested in compression 
($\dot{\epsilon}=1\!\times\!10^{-4}\,\mbox{s}^{-1})$ to measure the tensile-compressive asymmetry. 

Solution-treated and quenched samples were polished for examination using electron back-scattered diffraction (EBSD) on a field emission scanning electron microscope. 
A sample of Mg-3at.\%Zn, extruded under the same conditions, was also analysed in order to provide a comparison.
Pole figures and inverse pole figures were generated using OIM Analysis software, version 6.1.

The \betap precipitate size and spacing  were determined in the  peak-aged condition.
Samples for TEM analysis were prepared by grinding to \(\sim\!70\,\micron\) and thinning to perforation using a Gatan precision ion polishing system. 
TEM observations were conducted using  JEOL 4000EX and 2100 instruments operating at 400\,kV and 200\,kV, respectively.  
Stereological measurements utilised ImageJ software version 1.44. 
The cross-sectional areas of individual precipitates were determined from micrographs recorded along the  [0001] zone axis of the matrix. 
Precipitate cross-sectional areas were converted into an equivalent diameter.
The average centre-centre distance between \betap particles was determined by Delaunay triangulation \cite{Lepinoux2000}.
The mean interparticle spacing was determined by subtracting the mean precipitate diameter from the mean centre-centre distance.

TEM observations were also performed on a foil prepared from a region close to the fracture surface of a tensile specimen that had been pre-strained 5\% strain in tension, aged to peak hardness and then strained to failure.

\section{Results}

\subsection{Light microscopy}

The extent of twinning after pre-ageing deformation was measured by point counting (minimum of 500 points) with $3\pm1\%$ twins by volume found in both solution-treated and 3-5\% strained conditions (prior to isothermal ageing).
An average grain size of $17\pm3\,\mu$m was measured using line intercepts.

\subsection{Texture analysis}
\begin{figure*}
\begin{center}
\hfill
\subfigure[Mg-Zn-Y\label{fig-ebsd-ipf-1}]{\includegraphics[width=0.45\textwidth]{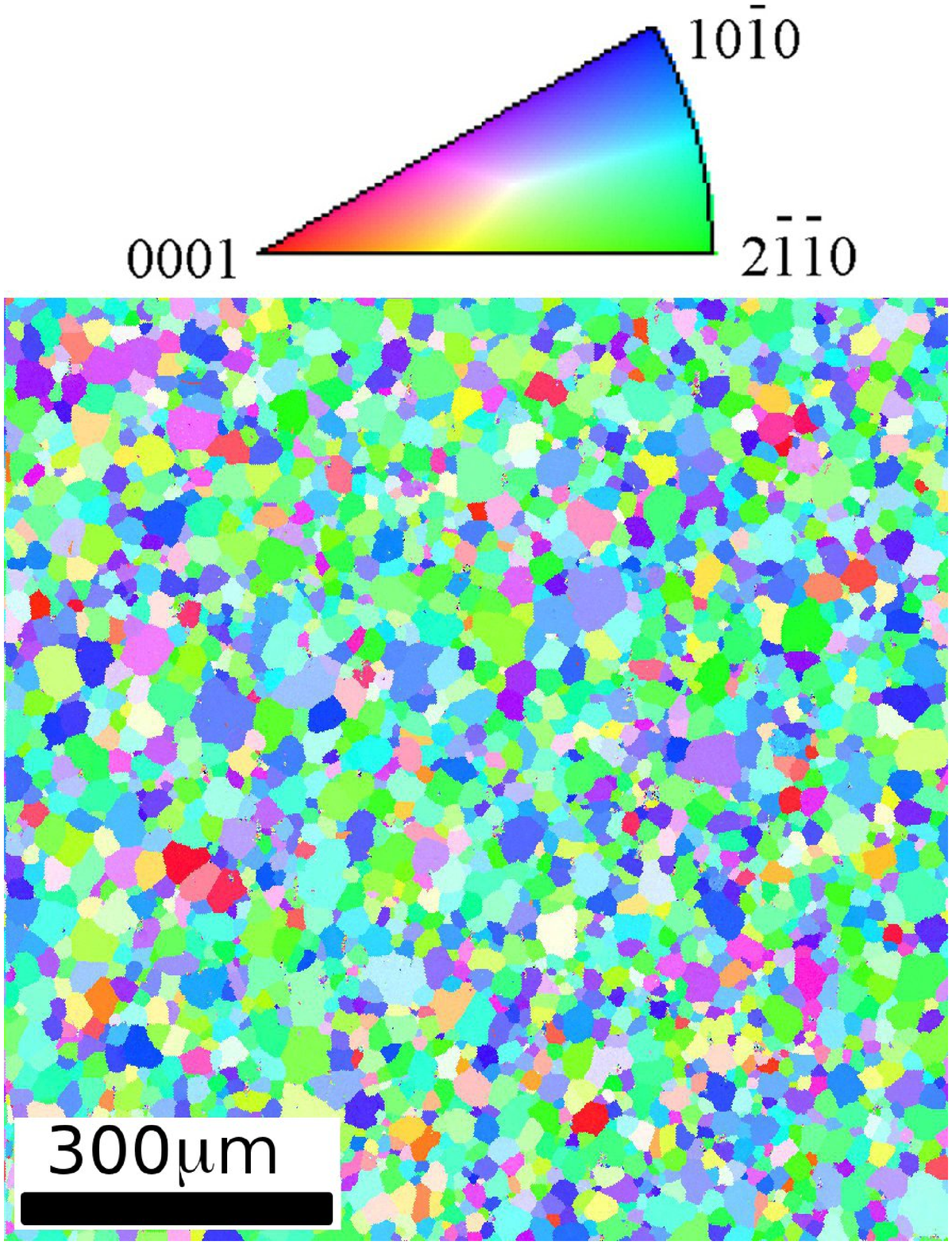}}\hfill
\subfigure[Mg-Zn\label{fig-ebsd-ipf-2}]{\includegraphics[width=0.45\textwidth]{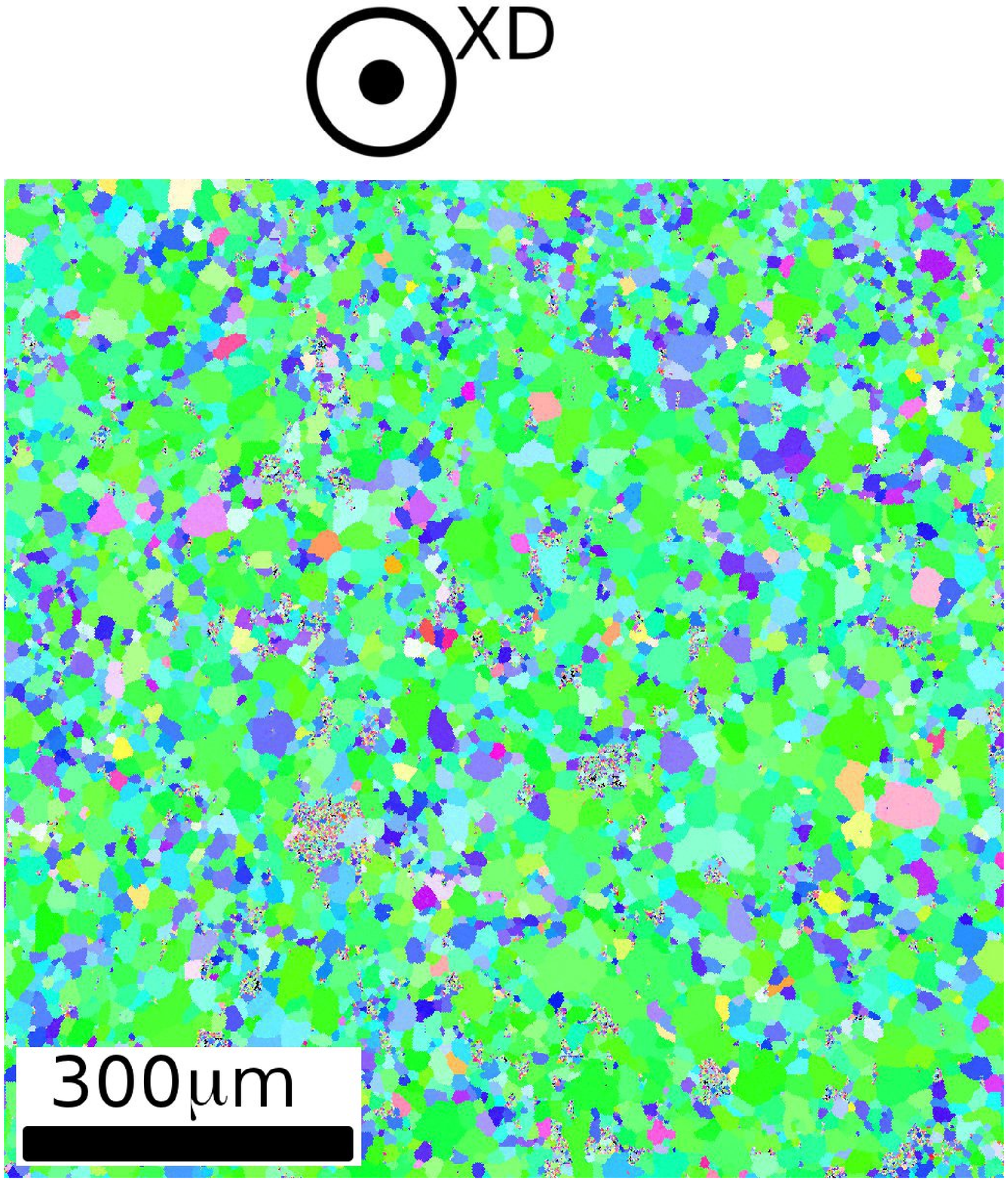}}\hfill\

\hfill
\subfigure[Mg-Zn-Y\label{fig-ebsd-pf}]{\includegraphics[height=3cm]{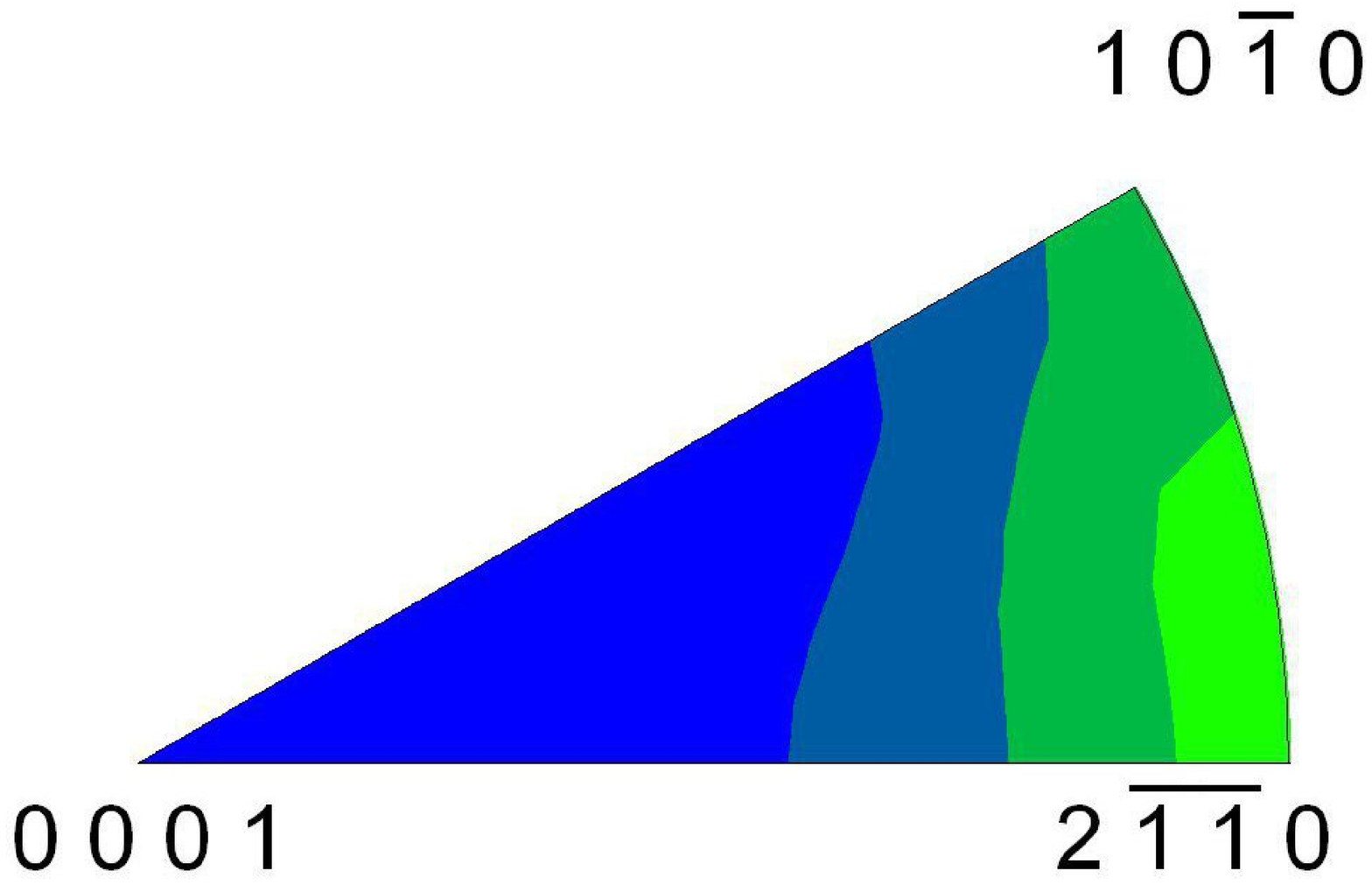}}\hfill
\subfigure[Mg-Zn\label{fig-ebsd-pf}]{\includegraphics[height=3cm]{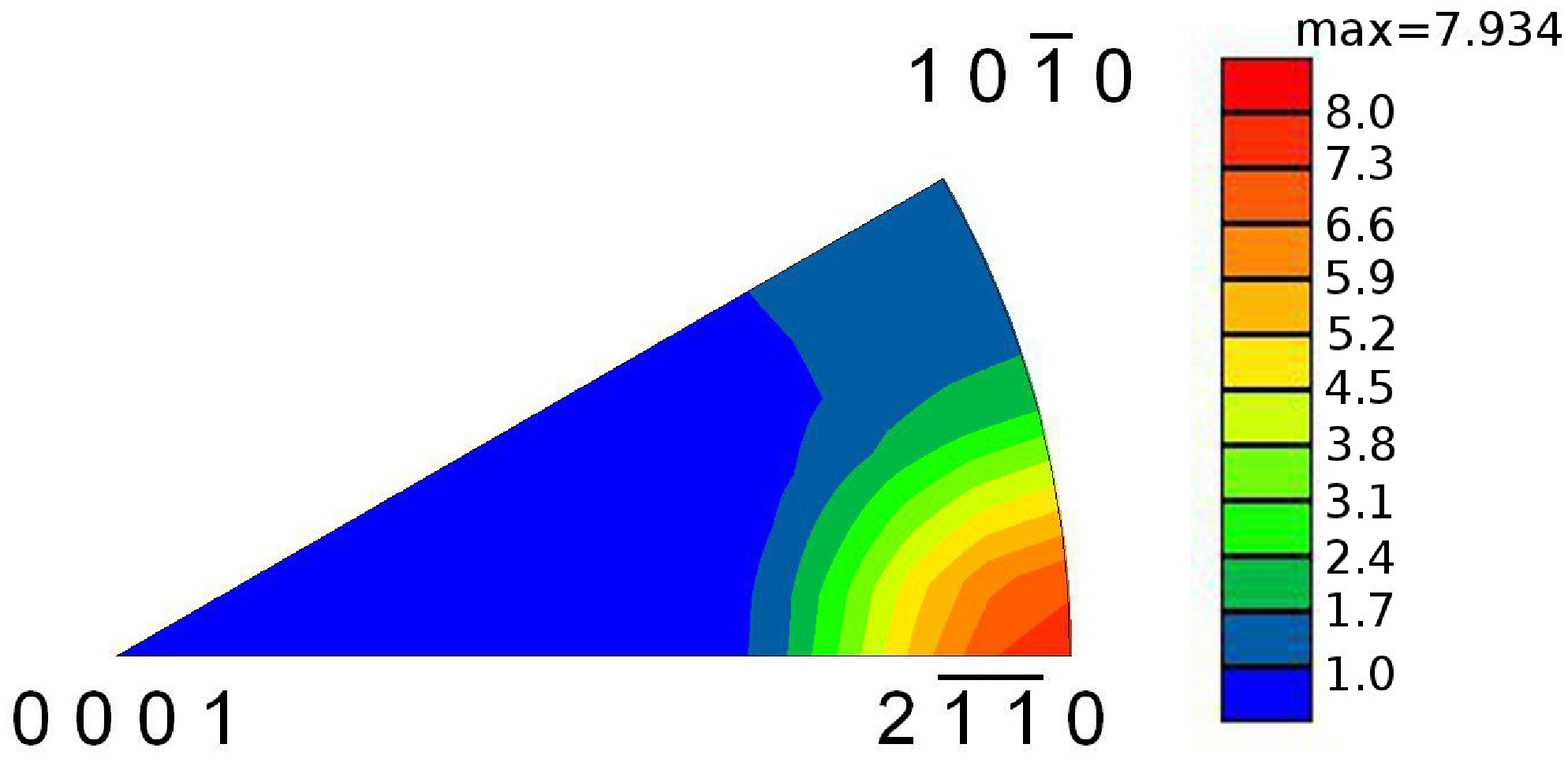}}\hfill\

\caption{Inverse pole figures and pole figures for  (a,c) extruded, solution-treated and quenched Mg-Zn-Y alloy and (b,d) a Mg-3at.\%Zn alloy extruded under the same conditions. 
The inverse pole figures (a,b) show a much weaker texture in the the Mg-Zn-Y alloy than in binary Mg-Zn.
The pole figures indicate that the nature of the texture is similar in each case, with  a predominance of grains with $\{2\overline{11}0\}$ plane normals parallel to the extrusion direction.
\label{fig-ebsd}}
\end{center}
\end{figure*}

The texture of cross-sections of extruded  Mg-Zn-Y were analysed using electron back-scattered diffraction (EBSD).
Samples of Mg-3at.\% extruded under similar conditions were also analysed. 
Inverse pole figures  showed  that for both alloys the prevalent orientation has the $(11\overline{2}0)$ plane normals close to the extrusion direction (XD) (Fig~\ref{fig-ebsd}(a,b)). 
Pole figures (Fig~\ref{fig-ebsd}(c,d)) showed that the intensity of this texture was much weaker in Mg-Zn-Y (c) where the maximum intensity was only 2.676, compared to Mg-Zn where the maximum was 7.934 (d). 
This corresponds to a higher fraction of grains in Mg-Zn-Y that have $(10\overline{1}0)$ planes aligned close to XD.

\subsection{Ageing response}

\begin{figure}
\begin{center}
\fig[0.475\textwidth]{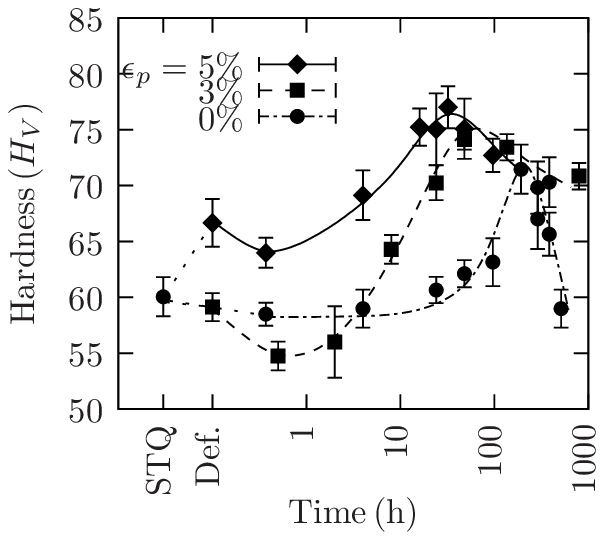}
\caption{Ageing response for the Mg-Zn-Y alloy at 150\celsius for various levels of  pre-ageing deformation. 
``STQ''  indicates the hardness in the solution treated and quenched condition.
``Def.'' shows the hardness after deformation (before ageing commenced). 
Error bars indicate the standard deviation in the data. 
 \label{fig-hardness}}
\end{center}
\end{figure}

Pre-ageing deformation substantially accelerated and enhanced the precipitation-hardening response of the Mg-Zn-Y alloy. 
Figure~\ref{fig-hardness} plots the hardness response for the alloy with various levels of prior plastic strain, $\epsilon_p$. 
The unstrained (T6 treated) alloy displayed a broad hardness plateau at approximately 60\,$H_V$  until 100\,h of ageing. 
The ageing response of unstrained (T6 treated) sample was delayed relative to that of a Mg-3at.\%Zn alloy, requiring 
256\,h to reach peak hardness compared to 48\,h for Mg-3at.\%Zn \cite{RosalieMgZn2012}.
This is consistent with a  previous report which noted that rare earth additions reduced the rate of overageing in Mg-Zn-RE alloys \cite{WeiPrec1995}.  

Ageing of T8 heat treated samples was more rapid, as expected, with 48\,h ageing required to reach peak hardness after 3\% pre-strain and 32\,h after 5\% pre-strain. 
The maximum hardness also increased by approximately 7\% from 72$\pm2$\,H$_V$ to 77$\pm2$\,H$_V$ for 5\% strain. 
Both the strained samples show similar shaped ageing curves, with the level of hardness in the case of 5\% strained sample  higher (77$\pm2$\,H$_V$) than that of 3\% strained sample (74$\pm1$\,H$_V$). 
In both cases, there was an immediate drop in hardness at the commencement of ageing, associated with annealing out of dislocations. 
In the case of 3\% strained sample this initially reduced the hardness to below that of unstrained sample, before the hardness increased again due to precipitation.

\subsection{Precipitate size distribution}

\begin{figure}
\begin{center}
\hfill
\subfigure[0\% strain\label{fig-3005ZY-0-side}]{\fig[0.4\textwidth]{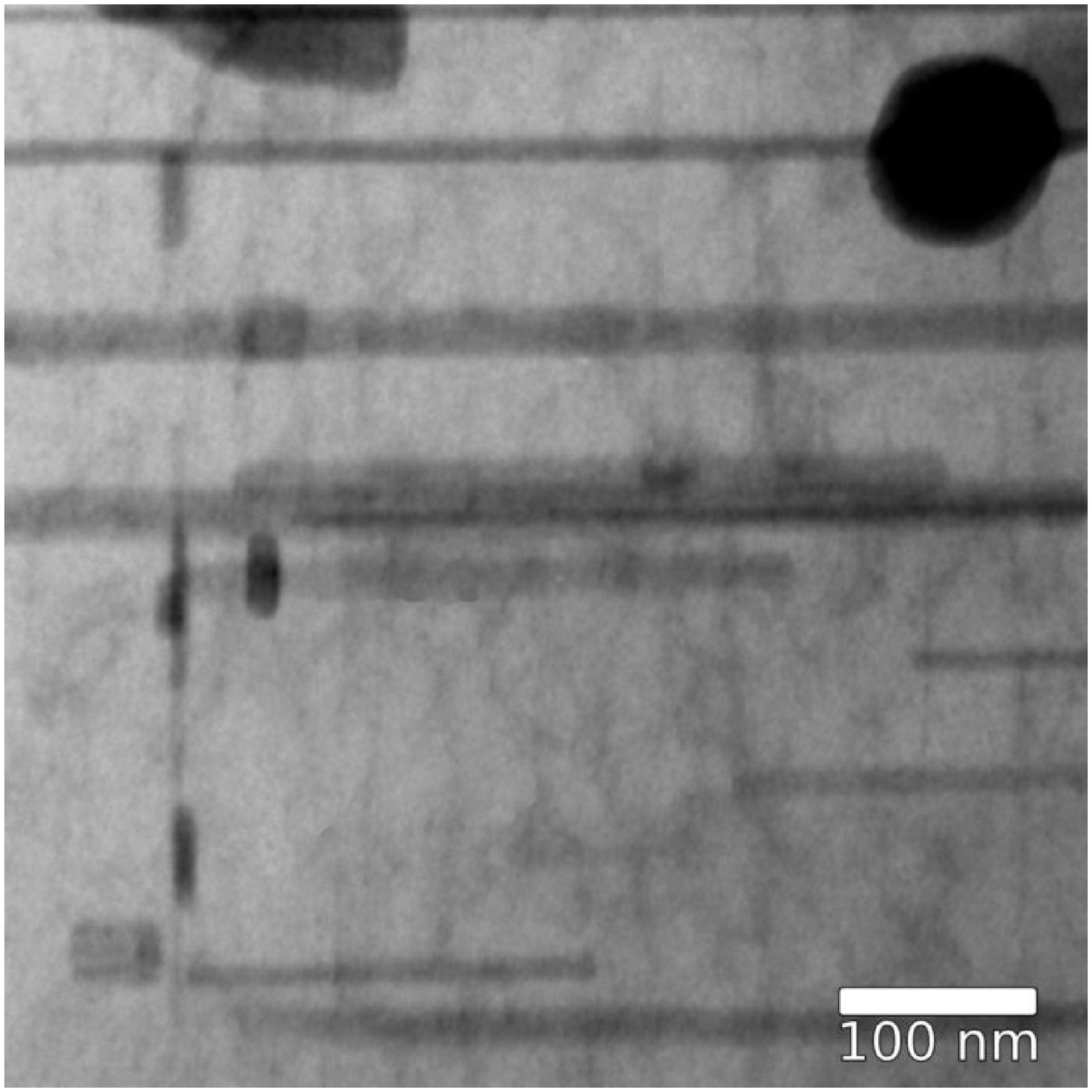}}
\hfill
\subfigure[3\% strain \label{fig-3005ZY-3-side}]{\fig[0.4\textwidth]{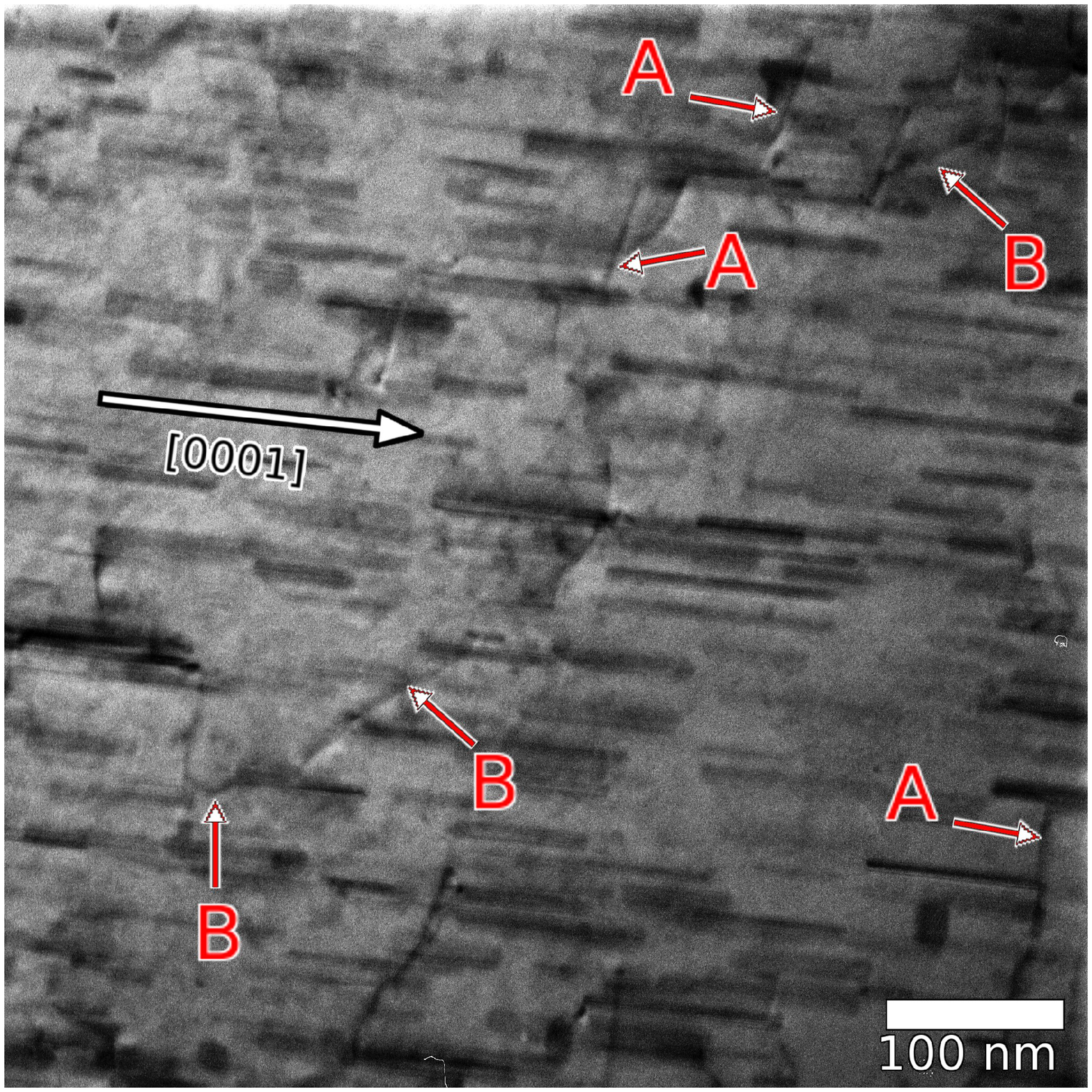}}
\hfill\

\subfigure[5\% strain \label{fig-3005ZY-5-side}]{\fig[0.4\textwidth]{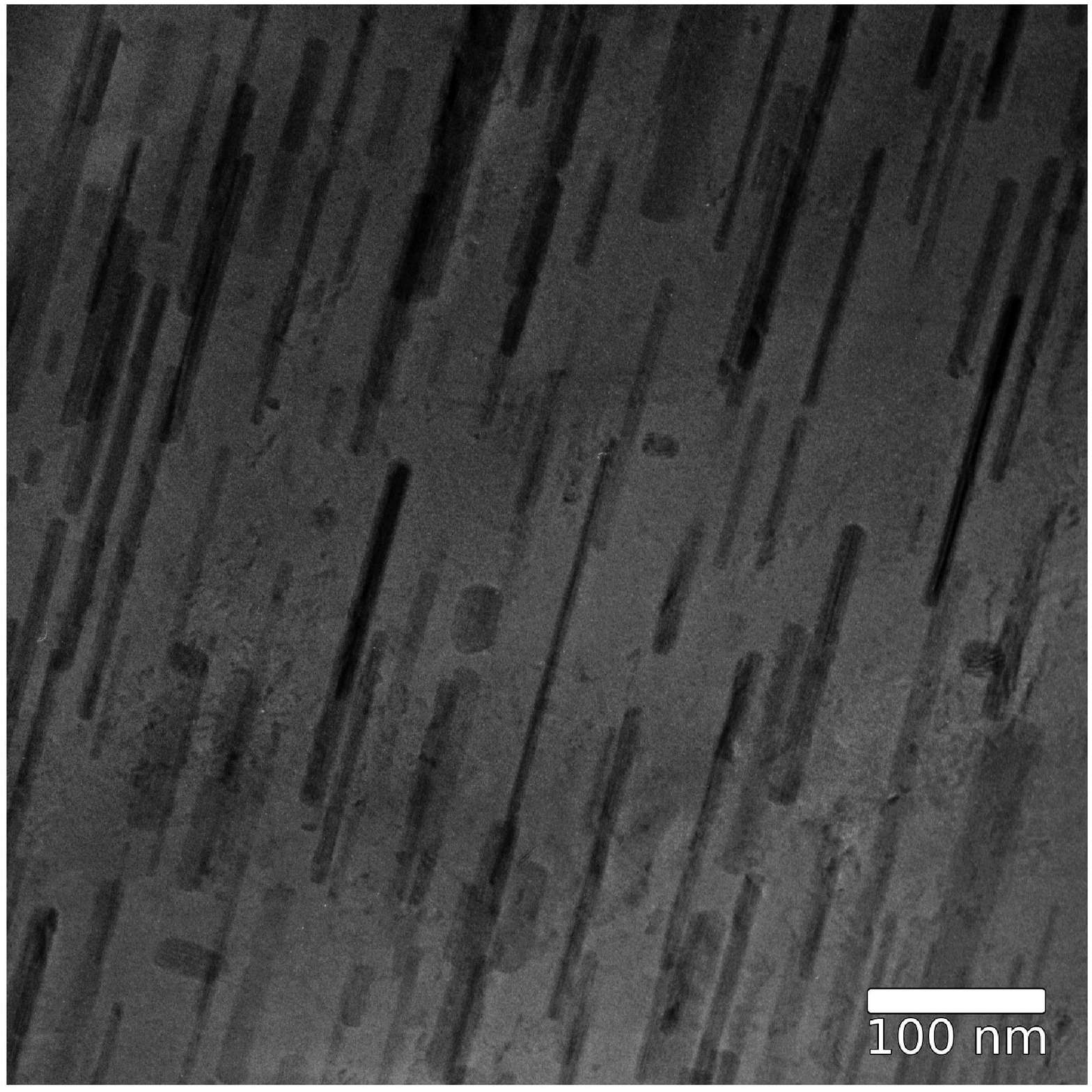}}

\caption{Transmission electron micrographs of \betap rod-shaped precipitates in the peak aged condition, showing \betap precipitates viewed normal to the hexagonal axis of magnesium.  
Dislocations are visible in the foil given 3\% strain (b).  
Segments of dislocations that are parallel to the basal plane are labelled ``A'' and segments of dislocations that lie at an angle to the basal plane are labelled ``B''.
\label{fig-tem-side}}
\end{center}
\end{figure}

\begin{figure}
\begin{center}
\fig[0.49\textwidth]{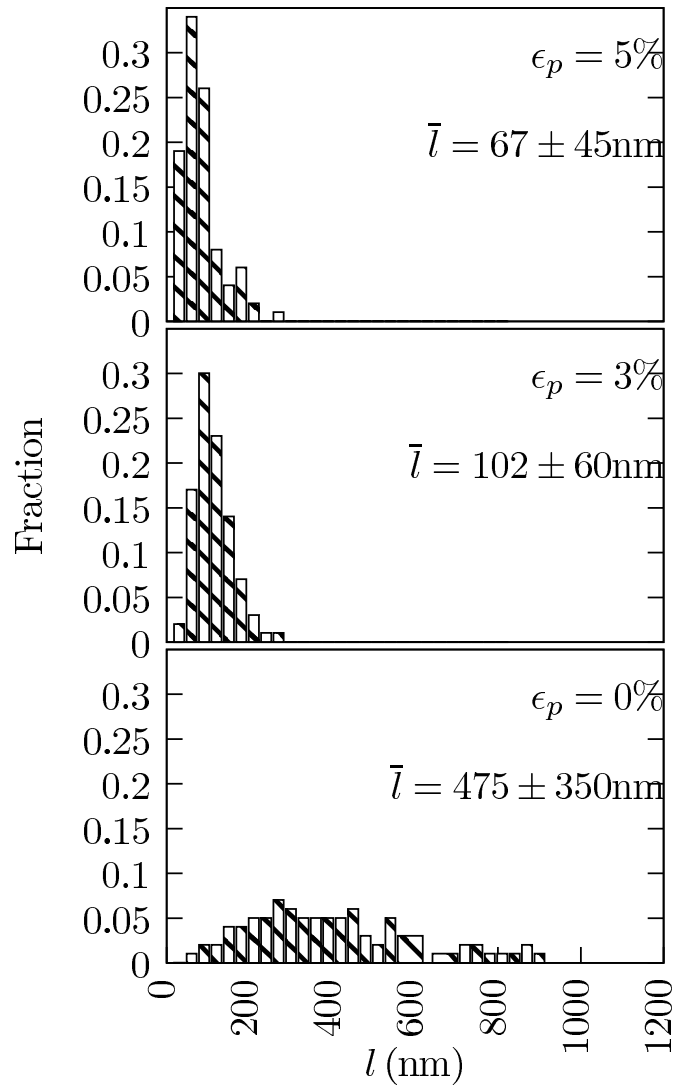}

\caption{Distribution of precipitate lengths of \betap precipitates in the peak aged condition for various amounts of  pre-ageing deformation.  
Deformation dramatically reduces the average precipitate length compared to non-deformed sample and results in a narrower distribution of lengths. 
\label{fig-length}}
\end{center}
\end{figure}

Pre-ageing deformation refined the \betap precipitate length by a factor of 7 and diameter by a factor by 2,
 reducing the aspect ratio of the rods from approximately 24 to 7.  
TEM micrographs of alloys aged to peak hardness (Figure~\ref{fig-tem-side}) clearly show the reduction in \betap precipitate length with increasing pre-ageing deformation.
Dislocations are frequently observed  in pre-strained samples, as indicated in  Figure~\ref{fig-3005ZY-3-side}.
Segments of some dislocations appeared to be parallel to the basal plane (those marked A in the figure), while in other cases dislocations (marked B) were at an angle to the basal plane. 
The distribution of precipitate lengths is set out in Figure~\ref{fig-length}.
Non-strained sample (Fig~\ref{fig-3005ZY-0-side}) contained \betap precipitates with a average length ($\overline{l}$) of 475\,nm. 
$\overline{l}$ was reduced to 102\,nm for 3\% strain (Fig~\ref{fig-3005ZY-3-side}) and  67\,nm for 5\% strain  (Fig~\ref{fig-3005ZY-3-side}).

T6 treated samples also contained plate-shaped $\beta_2^\prime$ precipitates with basal plane habit that can be seen in Figure~\ref{fig-3005ZY-0-side}.
The large spheroidal precipitate also present in this micrograph is consistent with previous reports of $i$-phase particles in Mg-Zn-Y alloys \cite{SinghTMS2008}.
Such spheroidal precipitates were extremely rare in samples given T8 treatments. 

\begin{figure}
\begin{center}
\hfill
\subfigure[0\% strain \label{fig-3005ZY-0-cross}]{\fig[0.4\textwidth]{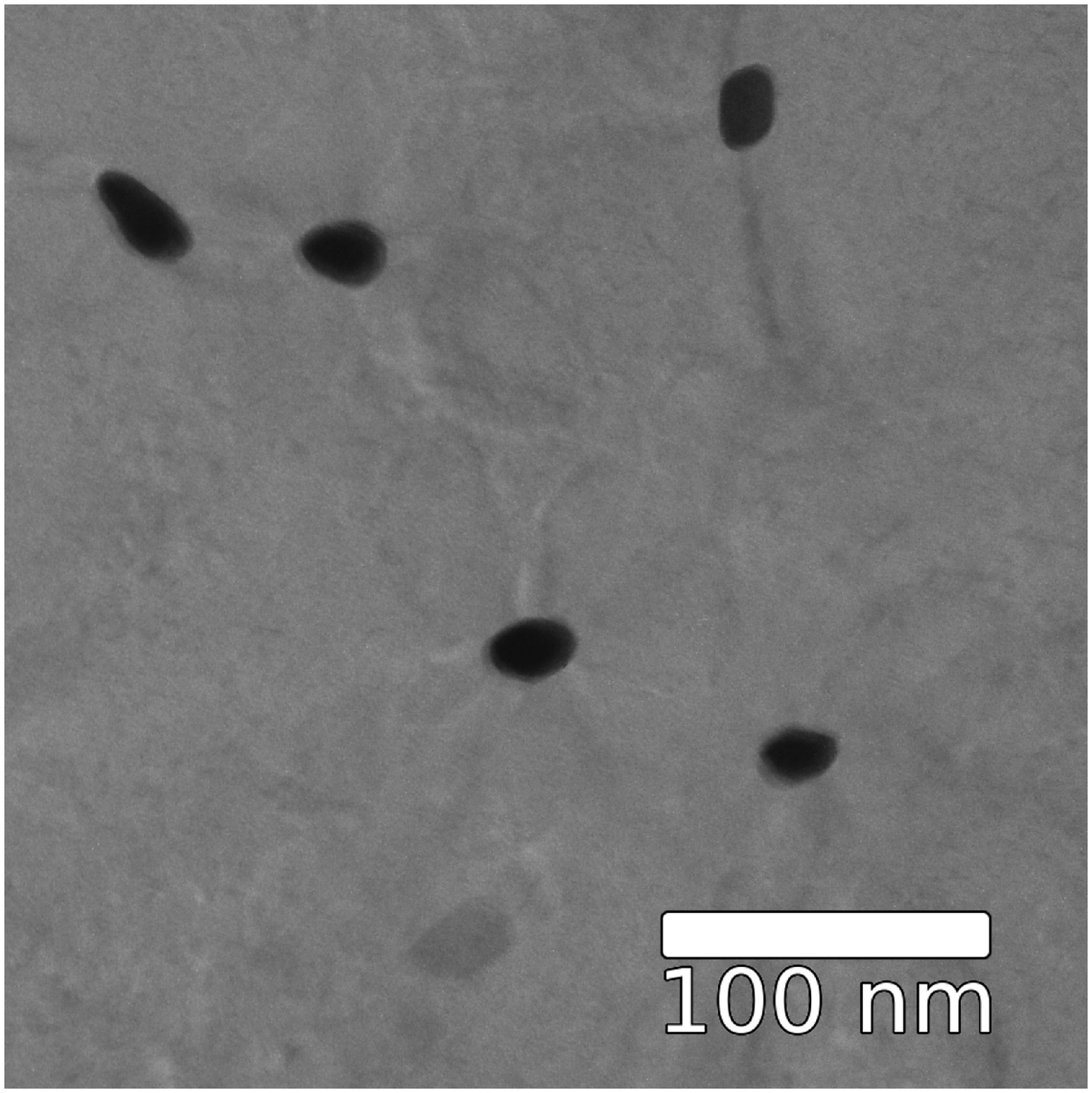}}
\hfill
\subfigure[3\% strain \label{fig-3005ZY-3-cross}]{\fig[0.4\textwidth]{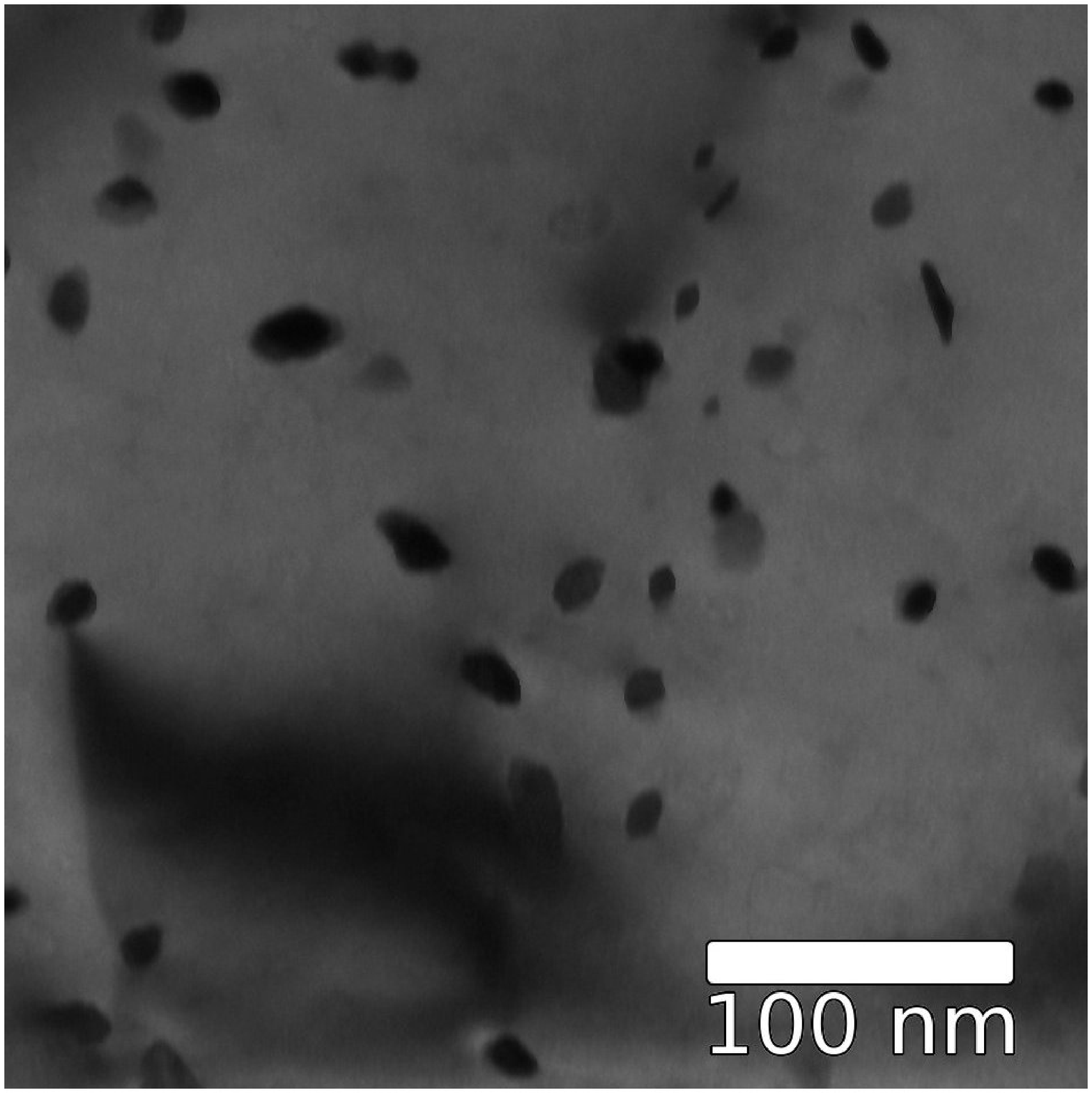}}
\hfill\

\subfigure[5\% strain \label{fig-3005ZY-5-cross}]{\fig[0.4\textwidth]{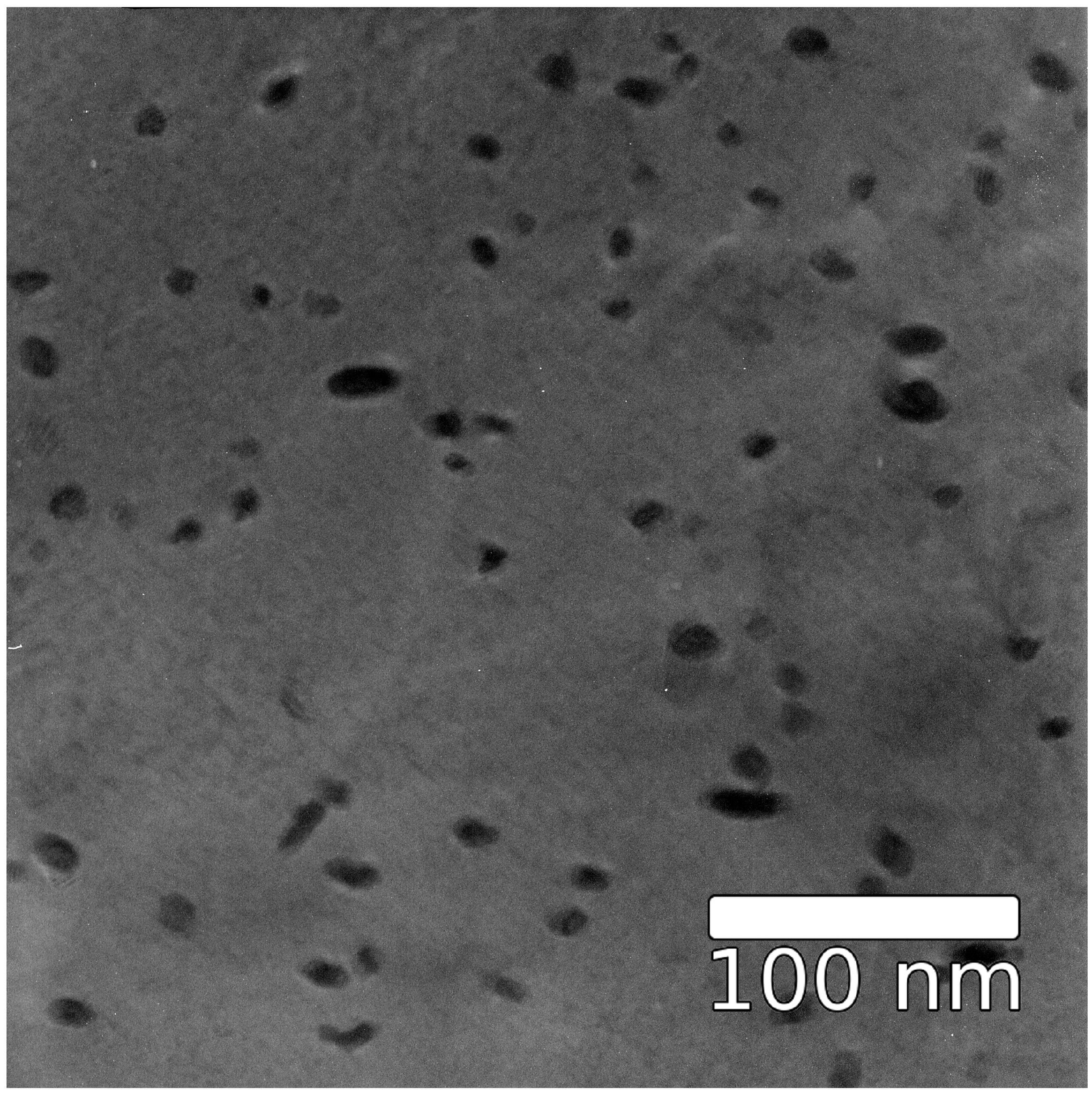}}

\caption{Transmission electron micrographs of \betap rod-shaped precipitates in the peak aged condition, showing \betap precipitates in cross-section; the electron beam was directed along the [0001] axis.
The foil in (a) was not deformed, while the foil shown in (b) experienced 3\% nominal strain prior to ageing and 
(c) was deformed to 5\% nominal strain before ageing. 
 \label{fig-tem-cross}}
\end{center}
\end{figure}

\begin{figure}
\begin{center}
\fig[0.49\textwidth]{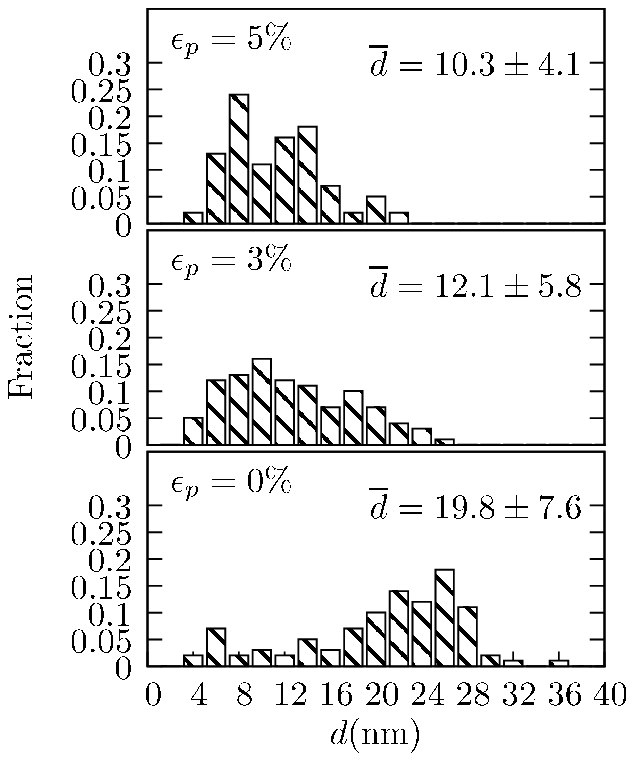}
\caption{Distribution of \betap precipitate diameters in the peak aged condition for various levels of pre-ageing deformation.
Samples deformed to 3\% or 5\% nominal strain show narrower distributions of particle diameters and lower average diameter values compared to the non-deformed samples.  \label{fig-diameter}}
\end{center}
\end{figure}

Pre-ageing deformation also reduced the diameter of the \betap precipitates.
Figure~\ref{fig-tem-cross} shows  TEM micrographs of alloys in the peak hardness condition, with the \betap precipitates in cross-section. 
Frequency plots of the precipitate diameter, $(d)$, are provided in Figure~\ref{fig-diameter}.
For 0\% pre-ageing strain (Fig.~\ref{fig-3005ZY-0-cross}) the precipitate distribution was sparse and relatively coarse, with  average diameter ($\overline{d}$) of 19.8\,nm. 
The precipitate diameters were relatively uniform, with 44\% of diameters falling in the range 22--26\,nm. 
When either 3\% (Fig~\ref{fig-3005ZY-3-cross}) or 5\% (Fig~\ref{fig-3005ZY-5-cross}) pre-ageing strain was applied over 50\% of  precipitates had $d\le12\,$nm. 
3\% tensile strain decreased  $\overline{d}$ to 12.1\,nm and this was further reduced to 10.3\,nm after 5\% strain. 
The precipitate spacing and number density are set out in Table~\ref{tab-lambda}.

%%%%%%%%%%%%%%%%%%%%%%%%%%%%%%%%%%%%%%%%%%%%%%%%%%%%%%%%%%%%%%%%%%%%%%%%%
\begin{table}
\begin{center}
\caption{Stereological measurements for \betap precipitates in alloys aged to peak hardness for various amounts of deformation. 
Values in parentheses indicate the standard error. 
$L_p$ is the centre-centre distance and $\lambda_b$ is the interparticle spacing, both of which were measured on the basal plane.
\label{tab-lambda}}

\begin{tabular}{p{3.5cm}r@{(}l@{ }r@{(}l@{ }r@{(}l@{ }}
\toprule
Alloy 	&	\multicolumn{6}{c}{Mg-3at.\%Zn-0.5at.\%Y}\\								\cmidrule(r){2-7}						
Deformation (\%)		&	\multicolumn{2}{c}{0}			&	\multicolumn{2}{c}{3}		&	\multicolumn{2}{c}{5}	\\
\raggedright Length\,(nm)	&	475	&	20)	&	102	&	4)	&	67	&	4)			\\
Diameter (nm)		&	20	&	0.6)	&	12	&	0.4)	&	10	&	0.2)			\\
$L_p$\,(nm)			&	298	&	5)	&	57	&	0.1)	&	56	&	0.2)			\\
$\lambda_b$\,(nm)	&	278	&	13)	&	62	&	2)	&	46	&	1)			\\
\raggedright Area fraction  (\%)* &	\multicolumn{1}{r}{0.7} &\multicolumn{1}{l}{} &	\multicolumn{1}{r}{3.9}		&\multicolumn{1}{l}{} &	\multicolumn{1}{r}{3.84} &\multicolumn{1}{l}{} \\

\raggedright Number density* \\$(\times10^{20}$ /m$^3)$	&	0.48	&	0.03)	&	33.3	&	2.5)	&	69 &	6.0)			\\
\bottomrule
\end{tabular}

\raggedright*Calculated (other values are for measured data)
\end{center}
\end{table}

\subsection{Tensile properties}

\begin{figure}
	\begin{center}
		\fig[0.45\textwidth]{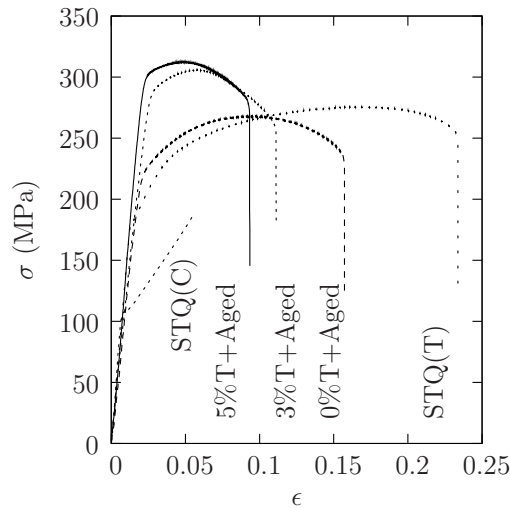}
	\caption{\label{fig-instron-eng}Stress-strain curves for various levels of pre-ageing deformation in tension.
``STQ'' indicates the solution-treated and quenched condition.
The initial portion of the curve for compressive deformation of this sample (labelled ``STQ(C)'') is also  included to show the existence of substantial asymmetry in the compressive and tensile yield strengths.}
	\end{center}
\end{figure}

Refinement of the precipitate microstructure by pre-ageing deformation increased the yield  strength  $(\sigma_y)$ with a corresponding reduction in tensile ductility.
Figure~\ref{fig-instron-eng} shows engineering stress-strain curves for the alloy with different levels of pre-ageing deformation.  
In the solution-treated and quenched (``STQ(T)'') condition $\sigma_y$  was 150\,MPa, which increased 
$\sim$50\% to  217\,MPa after ageing to the peak hardness condition.
The compressive yield strength of the solution-treated sample  (``STQ(C)'')  was $107\pm2$\,MPa 
and the asymmetry ratio $(\sigma_y(C)/\sigma_y(T))$ was 0.71, 
consistent with the asymmetry expected for an extruded Mg alloy with strong texture. 
Pre-ageing deformation (3\% strain)  increased $\sigma_y$   to 281\,MPa in tension, with little  further improvement for 5\% strain (287\,MPa).
The ultimate tensile strength (UTS) increased from $281\pm1$\,MPa (0\% strain) to  $322\pm2$\,MPa (5\% strain). 
However, if area-reduction is taken into account, (i.e. in terms of the true stress, where 
$\sigma_t=\sigma(1+\epsilon)$) the stress at failure  was $325\pm1$\,MPa for all samples except those aged without pre-strain, where it reached only $301\pm4$\,MPa.
 
The maximum uniform strain was determined from the  Consid\'ere criterion, i.e. the true strain at which $\sigma=\partial \sigma/ \partial \epsilon$ designated $\epsilon_c$. 
The true strain at fracture was determined by measuring the area reduction ($\epsilon_{ar}$) of fractured specimens.

\subsection{Post-fracture observation}
	
\begin{figure}
\begin{center}
\subfigure[Bright field, \textbf{B}={[}0001{]},  \textbf{g}= $10\overline{1}0$. \label{fig-post-fracture-a}\label{fig-post-fracture-a}]{\fig[0.48\textwidth]{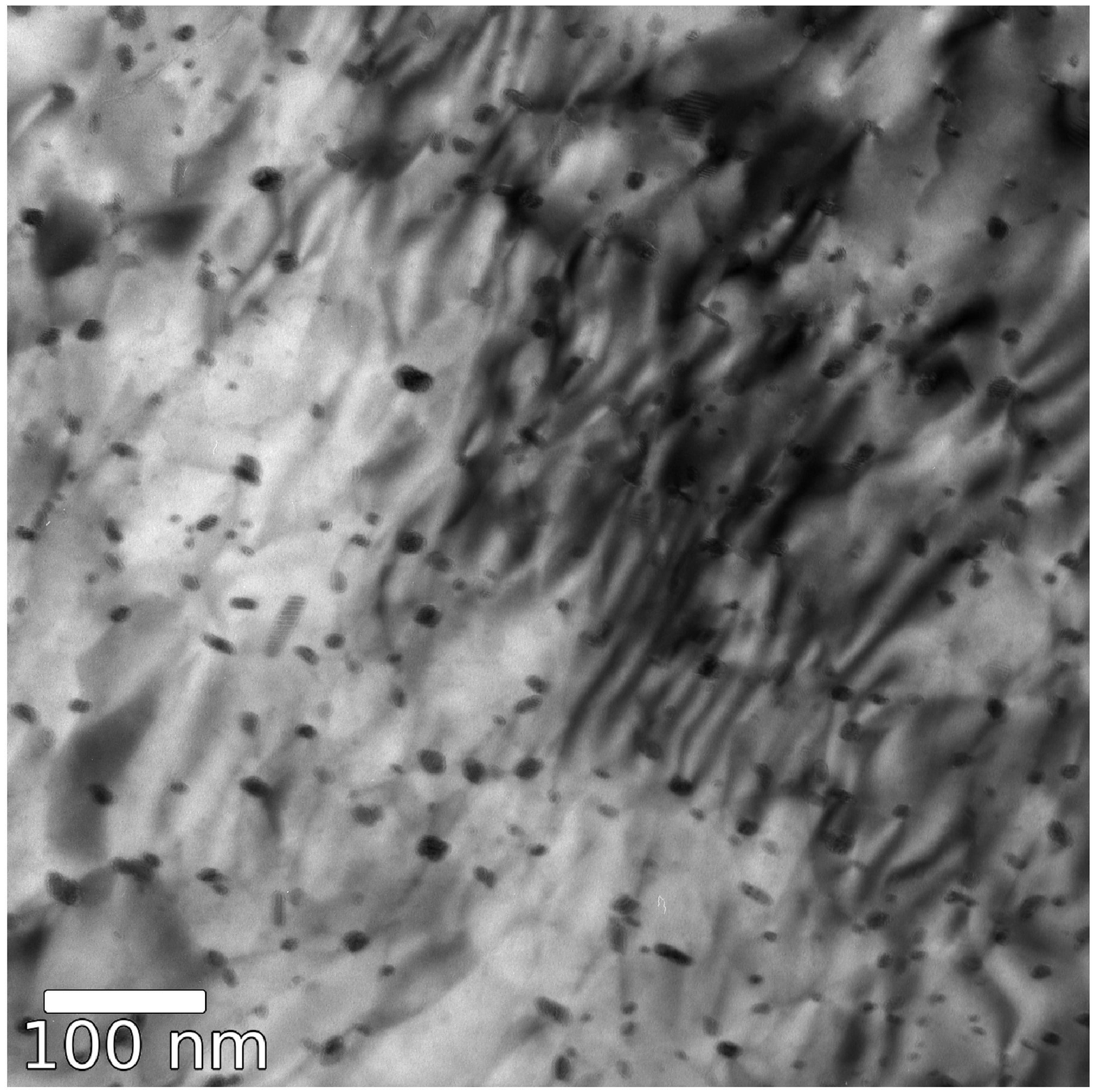}}\hfill
\subfigure[Weak beam, \textbf{B}={[}0001{]}, \textbf{g}= $10\overline{1}0$.\label{fig-post-fracture-b} \label{fig-post-fracture- b}]{\fig[0.48\textwidth]{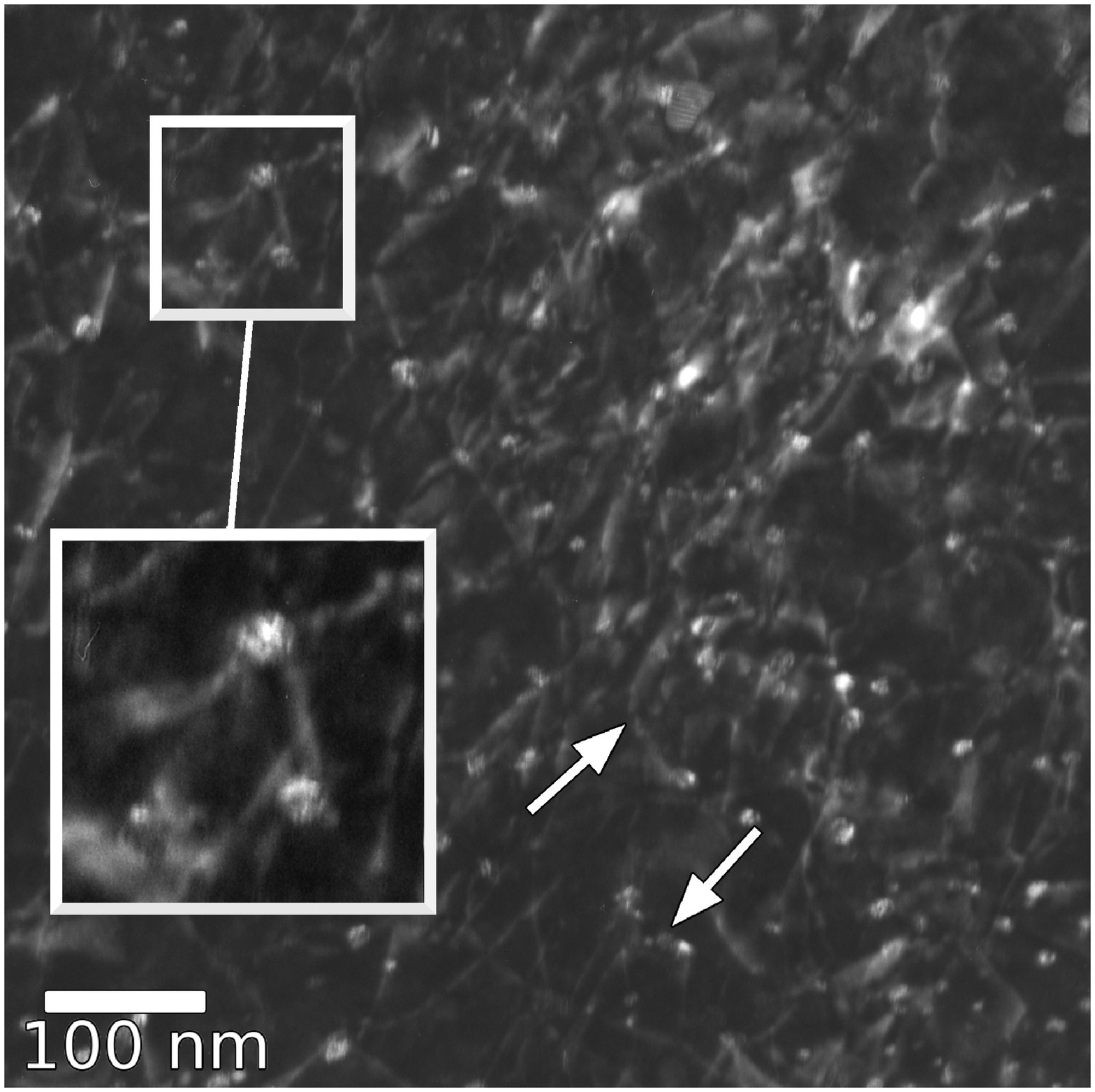}}

\subfigure[Bright field, \textbf{B}$\perp${[}0001{]}, \textbf{g}= $1\overline{2}10$.\label{fig-post-fracture-c}]{\fig[0.48\textwidth]{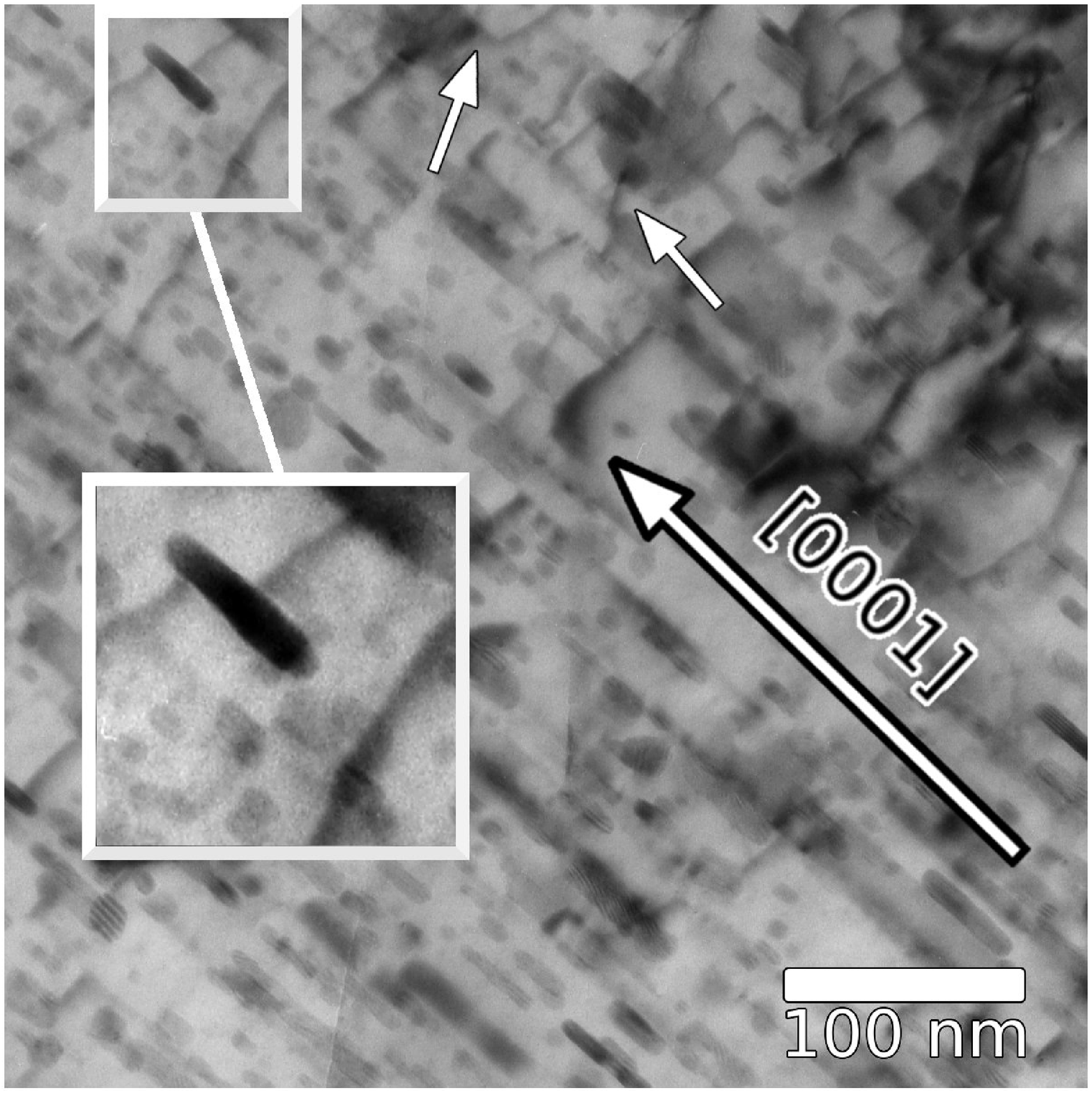}}
\caption{TEM micrographs from a fractured sample of Mg-Zn-Y. 
The sample was deformed to 5\% strain, aged to peak hardness and then strained to failure.
The foil was taken from close to the fracture tip. 
(a) and (b) show two-beam bright field and weak beam micrographs respectively, in a $\{10\overline{1}0\}$ diffraction condition with the electron beam close to the [0001] zone axis. 
Dislocations are readily apparent and frequently interact with the precipitates.
(c) shows a bright field micrograph with the electron beam normal to the [0001] axis. The enlarged inset shows a dislocation cross-slipping where it interacts with the precipitate.
\label{fig-post-fracture}}
\end{center}
\end{figure}

Foils obtained from regions close to the fracture surface of tensile samples were examined using TEM. 
Figure~\ref{fig-post-fracture} shows the microstructure of a sample given 5\% deformation, aged to peak hardness and deformed to failure. 
The sample was taken from close to the fracture tip. 
Fig~\ref{fig-post-fracture}(a,b) are a bright-field micrograph and a corresponding weak-beam micrograph  along the [0001] zone axis of a grain showing \betap precipitates in cross-section and dislocations in the basal plane. The bright-field micrograph is taken in a two-beam condition using a \{10$\overline1$0\} diffraction spot.
The weak beam image highlights the presence of dislocations which in most instances interact with one or more \betap precipitates. 
Many of these dislocations are bowed, indicating that the dislocations have additional line tension and suggesting that they are pinned by the \betap particles.
Two examples of such bowed dislocations are indicated by arrows in the figure and another is shown in greater detail in the  enlarged inset in Figure~\ref{fig-post-fracture-b}.
The inset show two \betap precipitates interacting with several dislocations, of which two are clearly bowed. 

Micrographs obtained with the beam normal to the [0001] axis of a grain (Fig~\ref{fig-post-fracture-c}) shows \betap precipitate rods side-on. 
Dislocation visible in the micrograph lie primarily in the basal plane.
Some dislocations appear to cross-slip when encountering a precipitate. Two such instances are indicated by arrows.  
Another instance is highlighted by enlarging in an inset, which shows a dislocation cross-slipping by interaction with a \betap precipitate. Such cases were frequently observed.
There is no indication of this precipitate being sheared by the dislocation.

\section{Discussion}

\subsection{Ageing behaviour \label{4-1}}

A 3\% pre-deformation caused no net increase in the hardness, and in the initial stage of ageing the hardness drops below that of the solutionised sample. 
This has presumably to do with diffusion of yttrium aided by dislocations, although further work is necessary to confirm this.
In case of a Mg-Zn alloy  \cite{RosalieMgZn2012}, in the initial stages of ageing, the hardness of 5\% pre-deformed sample at first dropped to the level of the 3\% deformed sample (which remained higher than that of unstrained sample) as the dislocations were annealed out and thereafter did not show any significant differences between these deformation conditions.  
In case of the ternary alloy in this study, the hardness of the 5\% pre-deformed sample remains higher than that of 3\% deformed sample at all stages of ageing. The difference in the hardness values is quite large in the beginning.

\subsection{Effect of pre-ageing deformation on  precipitate distribution}
\begin{figure}
\begin{center}
\fig[0.55\textwidth]{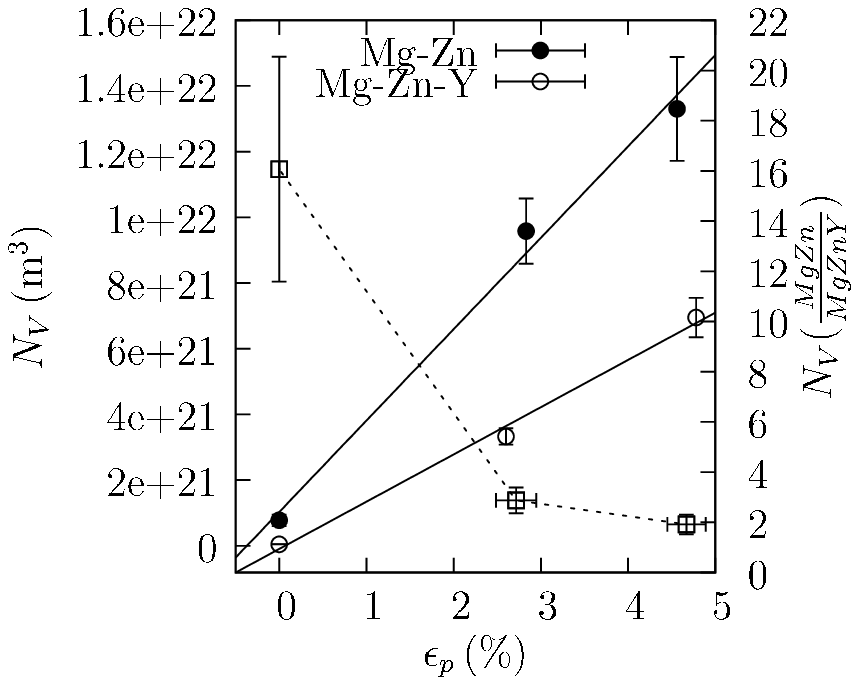}
\caption{The effect of pre-ageing deformation on the number density in samples aged to peak hardness for various amounts of pre-ageing strain ($\epsilon_p$).
The number density of \betap precipitates in the present alloy (open circles) was considerably lower than that in a similar
Mg-Zn alloy (filled circles) \cite{RosalieMgZn2012}. 
The ratio between the number density of the two alloys (squares, right axis) decreased with increasing pre-ageing deformation. 
  \label{fig-NumberDensity}}
\end{center}
\end{figure}

Pre-ageing deformation increased the number density of the \betap precipitates and reduced the precipitate length and diameter. 
Figure~\ref{fig-NumberDensity} plots the \betap number density in the peak-aged condition for Mg-Zn-Y (present work, open circles) and  a Mg-3at.\%Zn alloy \cite{RosalieMgZn2012} (filled circles) versus the pre-ageing deformation. 
In both cases there is a linear relationship between the plastic strain imposed on the alloy and the number density per unit volume for strains up to  5\%. 
The figure also displays the ratio of the \betap number density in the two alloys.
(the high uncertainty in the ratio of number densities in the unstrained samples was due to the low number density and higher error in this condition). 
Error bars are included in pre-ageing strain values to allow for slight differences in the plastic strains in the two alloys.
\betap number densities were greater in Mg-Zn alloys, however, pre-ageing reduces the ratio between the the two alloys from a factor of 16 (0\% strain) to $\sim$2 (5\% strain).

\begin{figure}
\begin{center}
\fig[0.55\textwidth]{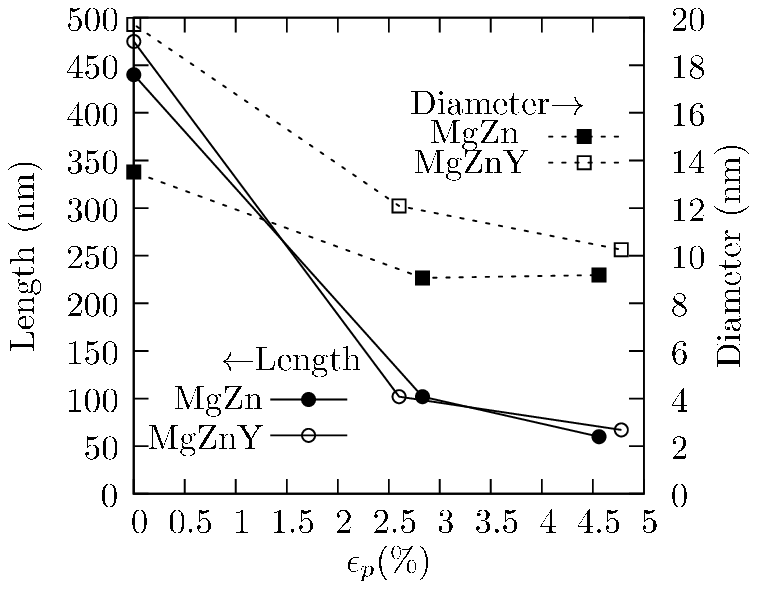}
\caption{The effect of pre-ageing deformation on the average precipitate length and diameter in Mg-Zn and Mg-Zn-Y. 
The deformation is given as the applied plastic strain, $\epsilon_p$.
Circles indicate average length values, while square represent diameter values. 
Open and filled symbols are for Mg-Zn-Y and Mg-Zn alloys respectively.  \label{fig-SizeCompare}}
\end{center}
\end{figure}

There was little difference ($\le11\%$) between the average \betap precipitate lengths in peak aged Mg-Zn-Y and those reported previously for Mg-Zn.
The precipitate diameter, however,  was up to 45\% greater in Mg-Zn-Y.  
Figure~\ref{fig-SizeCompare} compares the average precipitate length and diameter for Mg-Zn \cite{RosalieMgZn2012} and Mg-Zn-Y. 
In the non-deformed sample the average diameter in Mg-Zn-Y  was substantially greater than that observed in in Mg-Zn (19.7\,nm \textit{vs.} 13.5\,nm), but with 5\% pre-ageing strain the difference in  average diameter between the two alloys had narrowed to 10.3\,nm in Mg-Zn-Y \textit{vs.} 9.2\,nm in Mg-Zn.

Pre-ageing strain increases the volume fraction of \betap precipitates from approximately 0.5\% (T6 treatment) to 2.3\% (T8 treatment with 3 or 5\% cold-work).  
These values are less than the estimated 3.5\%  volume fraction of \betap precipitates in Mg-3at.\%Zn alloy after either T6 or T8 treatment  \cite{RosalieMgZn2012}. 
Calculations were based on the number density and average precipitate length and diameter determined from electron micrographs.
Yttrium is poorly soluble at the ageing temperature and is expected to precipitate as grain boundary ternary phases, primarily the quasicrystalline i-phase (\ce{Mg3Zn6Y})  since the Zn:Y ratio is close to  to the 6:1 ratio of this phase. 
The precipitation of grain boundary ternary phases will reduce the amount of zinc available for  precipitation as \betap rods and would explain the low volume fraction of \betap precipitates in the T6 heat treated condition. 
The increase in the \betap precipitate volume fraction in T8 treated Mg-Zn-Y alloy demonstrates that by providing heterogeneous nucleation sites (in the form of dislocations), pre-ageing deformation was able to re-partition Zn to \betap precipitates.
The changes in the solute partitioning may also be responsible for the ageing behaviour described in Section~\ref{4-1}.

\subsection{Deformation mechanism}

Extrusion of the alloy results in a tube texture, with the basal plane normals aligned radially. 
Pole figures show a high concentration of basal planes in the radial direction, with a pair of prismatic planes aligned towards the extrusion direction (Fig~\ref{fig-ebsd}).  
Mg-Zn-Y has a relatively weak texture, with a maximum intensity of 2.676, compared to a maximum of 7.934 in a similarly extruded Mg-Zn alloy.
The weakness of this texture was thought to be due to the addition of yttrium, which has been shown to weaken the extrusion texture \cite{BallPrangnell1994,BohlenTexture2007,StanfordTexture2008}. 
This is thought to be due to the presence of high-melting temperate ternary phases which stimulate recrystallisation.
 
Tensile deformation of grains  with  $(11\overline{2}0)$ orientation provides c-axis compression. 
However, the low strain $\{10\overline{1}2\}$ twinning mode in Mg provides c-axis extension and therefore is not expected to contribute to deformation during pre-ageing strain. 
This is consistent with the low (3$\pm 1$\%) volume fraction of twins observed via light microscopy of deformed samples and  indicates that slip was the predominant deformation mode during pre-ageing deformation. 

The compressive-tensile asymmetry of solution-treated Mg-Zn-Y was due largely to twinning. 
Compression parallel to the extrusion axis provides c-axis tension and is likely to activate low-strain $\{10\overline{1}2\}$ twinning at lower stresses than either basal or prismatic slip. 
The compressive yield strength of the gives an asymmetry ratio $(\sigma_y(C)/\sigma_y(T))$ of 0.71.
This was close to the value of 0.75 measured in  Mg-3at.\%Zn \cite{RosalieMgZn2012} but greater than the value of  0.61 reported for a Mg-5wt.\%Zn alloy aged without deformation \cite{Robson2011}.
It has been reported that precipitates such as \betap may alter the compressive-tensile asymmetry of heat treated samples \cite{Robson2011} and further work to verify this is underway.

Dislocation glide in Mg alloys can involve slip on basal, prismatic or prismatic planes. 
Calculations of the Schmid factor from the EBSD data suggest a higher average Schmid factor for basal slip (0.33) than for prismatic (0.28), but there was considerable scatter in the values, with a standard deviation of 0.14 in both values. 
This would suggest that there may be a slight preference for basal slip, however, both slip systems are likely to be active during pre-ageing deformation. 
This differs from the results of Al Samman \textit{et. al}'s study \cite{AlSamman2010}, where tensile deformation of an extruded AZ31 alloy occurred principally by prismatic rather than basal slip. 
Prismatic slip in Mg alloys have also been reported elsewhere (see, for example \cite{KoikeNonBasalSlip,KoikeNonBasal2003a}).  
(Although the Schmid factor for $\langle c+a \rangle$ slip was non-zero, the critical resolved shear stress is high  \cite{Raeisinia2010} and this deformation mode was not expected to be significant).

There was little evidence to suggest that the precipitates were sheared by dislocations. 
The mechanical properties were not consistent with those expected for alloys containing shearable particles which have a tendency for coarse, planar slip \cite{Blankenship1993} and poor work hardening.
In contrast, the stress-strain curves for Mg-Zn-Y (Fig~\ref{fig-instron-eng}) show work-hardening behaviour in T6 and T8 conditions. 
This would be expected for non-shearable precipitates where dislocation pile-ups result in an additional back-stress on nearby dislocations. 
In addition, there was no correlation between the yield strength increase ($\Delta \sigma_y$) and  equations based on Friedel's effect \cite{FriedelDislocations1964} where
\( \Delta \sigma_y = c f^a r^b
\)
and $f$ is the precipitate volume fraction, 
$r$ is the particle radius and 
$a,b$ and $c$ are empirical constants, with $a,b=0.5$ \cite{shercliff:1990a}.

Post-fracture transmission electron microscopy did not  provide conclusive evidence of whether the \betap were sheared by dislocations. 
However, \betap precipitates of similar size to the peak-aged condition were present in foils taken from fractured samples.
Dislocations lying between these particles were frequently bowed, as would be expected for dislocations pinned by non-shearable particles (Fig~\ref{fig-post-fracture-b}).
Previous studies on compressive deformation of aged  Mg-5at.\%Zn alloys also reported that the precipitates were not sheared by either slip or twinning \cite{StanfordBarnett2009}. 
This is consistent with the poor coherency between the \ce{MgZn2} and \ce{Mg4Zn7} crystal structure and the matrix and also with the high elastic modulus of the intermetallic precipitates (estimated at $G$, 32-35\,GPa for \ce{MgZn2}  compared to  20.23--20.38\,GPa for Mg \cite{WuLaves2010}.)

The effectiveness of rod-shaped precipitates in impeding slip is dependent on the slip system \cite{Robson2011} and it is important to consider which slip system is active when peak-aged samples are deformed. 
The average Schmid factors for basal and prismatic slip are close enough to suggest that slip may occur by either mechanism, depending on the alignment of the individual grains. 
Transmission electron micrographs of the deformed and aged Mg-Zn-Y alloys show dislocations with segments lying in the basal plane and also crossing between basal and non-basal planes, as noted in Figures~\ref{fig-3005ZY-3-side}.
The post-fracture micrograph (Fig~\ref{fig-post-fracture-c}) also showed dislocations on the basal plane that appear to have cross-slipped along the precipitate before continuing along the basal plane (Fig~\ref{fig-post-fracture-c}).
Micrographs of similarly deformed Mg-Zn \cite{RosalieMgZn2012} also showed evidence that would suggest both basal and prismatic slip were active. 
Since both basal and prismatic glide involve dislocations with Burgers vector $\langle a \rangle$,  cross-slip can occur under the influence of stress, or in response to impassable obstacles, such as the \betap precipitates.  
It was considered that cross-slip between basal and prismatic planes may well occur. 
Therefore, both slip systems were taken into account in determining the effect precipitate size and distribution on the mechanical properties.

\subsubsection{Basal slip}

The average spacing between particles on the basal plane could be calculated directly from TEM micrographs using Delaunay triangulation \cite{Lepinoux2000} (see Table~\ref{tab-lambda}). 
This method has been shown to give a  representative measure of the effective particle spacing \cite{Lepinoux2000, Lepinoux2001} and does not rely on the assumption that the distribution is homogeneous.

\begin{comment}
\begin{figure}
\begin{center}
\fig[0.55\textwidth]{LambdaCompare}
\caption{Measured and calculated interparticle spacings on the basal plane. 
Filled/open symbols are for Mg-Zn and Mg-Zn-Y, respectively. 
The solid line shows equal measured and calculated spacings. 
A line of best fit though the data (dashed) shows that the distribution is inhomogeneous, with 
$\lambda\mbox{(Meas.)}>\lambda\mbox{(Calc)}$ in all cases.
  \label{fig-LambdaCompare}}
\end{center}
\end{figure}

Figure~\ref{fig-LambdaCompare} plots the measured interparticle spacings on the basal plane against values  
calculated via the equation for basal slip around rod-like precipitates.  \cite{NieMg2003}: 
\begin{equation}
 \lambda = \left( \frac{0.953}{\sqrt{f}} -1 \right) d_1 
\label{eq-lambda}
\end{equation}
The plot shows measured and calculated values for Mg-Zn-Y (open squares) and Mg-Zn\cite{RosalieMgZn2012}(filled squares). 
All the data falls below the solid line (which indicates equal calculated and measured values), showing that the equation underestimates the interparticle spacing by 10-15\% in both Mg-Zn-Y and Mg-Zn alloys. 
\end{comment}

\subsubsection{Prismatic slip}

Interparticle spacings were calculated using number density and precipitate length values measured from micrographs. 
The equation for prismatic slip proposed by Robson  \textit{et al.} modelled the precipitate distribution using a square grid, finding the relationship  \cite{Robson2011}: 
\begin{equation}
 \lambda = \frac{1}{\sqrt{N_A}} - l
\label{eq-lambda}
\end{equation}
with the number density on a single slip plane given by; $N_A=N_V d_t$ where $N_V$ is the number density per unit volume and $d_t$ is the precipitate diameter  and $l$ is the precipitate length.
However, placing the particles on a square grid means that the long axes of the rods have no lateral displacement from adjacent precipitates, making the spacing highly sensitive to the precipitate length. 
Indeed, $ \lim_{l \to N_A} (\lambda) = 0 $, implying that the precipitates form a continuous network.

The use of a triangular grid of precipitates  gives a more complex relationship between precipitate length and spacing, but avoids the unlikely situation that arises when rods are co-linear along the [0001] direction.
The centre-centre distance, $Lp$ is a function of the number density alone, with:
$ L_p = 1/\sqrt{N_A}$ 
and if the particle length $(l)$ is expressed as a multiple $(k)$ of $(L_p)$\cite{RosalieMgZn2012}.
\begin{equation}
\lambda			 =  \frac{1}{\sqrt{N_A}} \sqrt{ \left( (k^2+1 - k\sqrt{3} \right)} \label{pyramid}
\end{equation}
The equation still breaks down for high aspect ratio particles,  suggesting a minimum of $\lambda=0.5L_p$ for $l=L_p$ and \textit{increasing} $\lambda$ value for $l > L_p$. 
As a more accurate treatment is not yet available, calculations were based on the triangular grid, and  where $l > L_p$, the spacing used  was  $\lambda=0.5L_p$ (the minimum value from the Equation~\ref{pyramid} ).
Interparticle spacings for  prismatic plane spacing are listed in Table~\ref{tab-lambda-py}. 

\begin{table}
\begin{center}
\caption{Interparticle spacings on the  prismatic ($\lambda_p$) planes  and basal ($\lambda_b$) planes and the measured yield strength for Mg-Zn-Y. 
\label{tab-lambda-py}}
\begin{tabular}{lcrl@{(}c@{)}}
\toprule
Strain	& $\lambda_p$	&$\lambda_b$ &	\multicolumn{2}{p{5ex}}{$\sigma_y$} \\
		& (nm)		& (nm)	& \multicolumn{2}{p{5ex}}{(MPa)} \\
\midrule
STQ		&N/A 		&	N/A   &150&	1\\
0.00\%	&474 	&	278	&217&	2\\
3.00\%	&102		&	62	&281&	10\\
5.00\%	&67		&	46	&287&	10\\
\bottomrule
\end{tabular}
\end{center}
\end{table}

\subsection{Orowan looping}
Room temperature deformation of alloys containing non-deformable particles is regarded as involving dislocations bowing and eventually looping around particles in a process termed Orowan looping. 
The  increase in yield strength ($\Delta \sigma_y$) for non-shearable particles can be described by the well-known equation:
\begin{equation}
\label{eq-mg-orowan-rod}
\Delta\sigma_y =
 \frac{Gb}{2\pi\sqrt{1-\nu}}
\frac{1}{\lambda}
\ln{\frac{d_t}{b}}
\end{equation}
\noindent where $\nu$ is Poisson's ratio, 
$G$ the shear modulus (GPa),
 $b$ is the magnitude of the Burgers vector for basal slip in Mg \cite{AvedesianMg99} 
and $\lambda$ is the average interparticle spacing.
Equations for the interparticle spacing have been developed for a range of precipitate geometries. 
The actual interparticle spacing in the ternary alloy will depend only slightly on the presence of spheroidal i-phase particles that precipitated during solidification and were dispersed through the matrix during extrusion. 
However, noting that the spheroidal particles are both limited in number and relatively coarse, their presence was neglected.

\begin{figure}
\begin{center}
\includegraphics{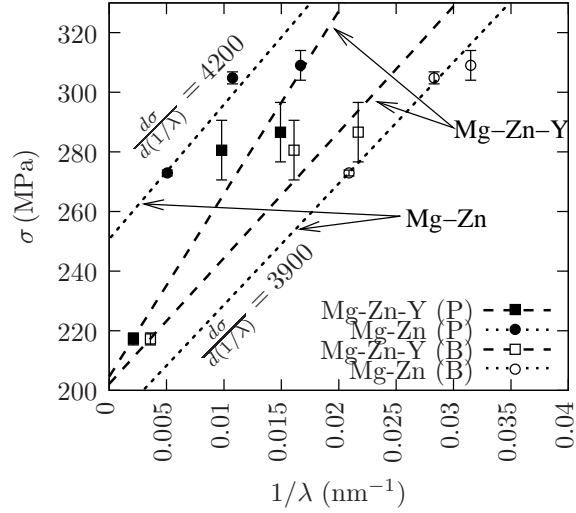}
\begin{spacing}{2}\caption{Yield strength ($\sigma$) for Mg-Zn-Y (present work) and Mg-Zn \cite{RosalieMgZn2012} alloys plotted against reciprocal of precipitate spacing  ($1/\lambda$) in the prismatic (P) and basal (B) planes.  \label{sigma}}
\end{spacing}
\end{center}
\end{figure}

Both Mg-Zn-Y and M-Zn alloys show a linear relationship between the yield strength and $1/\lambda$, with similar gradients for 
basal slip in both alloys and prismatic slip in Mg-Zn. 
Figure~\ref{sigma} plots the measured yield strength of the Mg-Zn-Y peak-aged alloy (squares) and Mg-Zn (\cite{RosalieMgZn2012} (circles) against the reciprocal of the measured interparticle spacing (1/$\lambda$) for \betap precipitates.
Filled and open symbols indicate the spacing on prismatic and basal planes, respectively.
The gradient  measured for prismatic slip in Mg-Zn-Y , however, is 6100\,MPa, greater than predicted by Orowan equation calculations (which predict 4760\,MPa).
This suggests that the interparticle spacing (and Orowan strengthening) is not well modelled for prismatic slip around low number densities of long ($>400$\,nm) precipitates. 
In such cases precipitate-dislocations interactions would be infrequent, but the particles would be extremely difficult to bypass. 
Extrapolation to $1/\lambda =0$ gives an estimate of yield strength of 202\,MPa for basal slip in Mg-Zn-Y, greater than the value of 187\,MPa for Mg-Zn. 
In both alloys the yield stress for prismatic slip is greater by $\sim$50\,MPa for equally spaced particles.
While the relative amounts of prismatic and basal slip are not accurately known, 
 it seems likely that, where possible, dislocations will cross-slip  to the basal plane where it appears that the  precipitates  offer less resistance to slip.

\subsection{Ductility}

The ductility of solution-treated Mg-Zn-Y was reduced after ageing to the peak-aged condition, with the reduction being more severe after T8 treatment where the interparticle spacing was reduced. 
The ductility of particle-containing systems has been examined by Chan \cite{chan1995}, whose model (since extended by Liu \textit{et. al} \cite{LiuZhang2004}) considers the build up of dislocations around precipitates, with failure occurring once a critical dislocation density is reached.  
In microstructures containing refined obstacles the sites of dislocation accumulation are closer together and the local dislocation density is higher. 

In the case where the \betap precipitates are the barriers to dislocation motion the geometric slip distance is the interparticle spacing, $\lambda$, on a given slip plane. 
The dislocation density ($\rho$)  is given by:
$\rho = (4 \epsilon /\lambda{b})  $
where $\epsilon$ is the  strain and $b$ is the magnitude of the Burgers vector for slip \cite{Ashby1970}. 
Failure is assumed to occur when the local dislocation density reaches a critical value, $\rho_{cr} $, which takes place at a local critical strain $\epsilon_{cr}$ given by:

\begin{equation}
\epsilon_{cr} = \frac{1}{4}\rho_{cr} b \lambda \label{eqn-critical-stress}
\end{equation}

A small correction is required to account for work-hardening behaviour \cite{RambergOsgood1943,Hutchinson1968}  using coefficients $I$ and $h$ \cite{LiuSun2005,Dowling1987} where :
\begin{eqnarray}
I &= &10.3 \sqrt{0.13+n}-4.8n\\
h &= &\frac{3}{2\sqrt{1+3n}}
\end{eqnarray}

With  work-hardening and localisation taken into account \cite{chan1995,LiuZhang2004} the macroscopic strain to failure, $\epsilon_f$, for cylindrical rod precipitate of length $l$ is given by:
\begin{equation}
\epsilon_f = 
\frac{1}{\tilde{\epsilon}_E(\theta)} 
 \left[
\frac{I}{0.405 \pi l }
\right]^ \frac{1}{n+1}
\frac{\tilde{\epsilon_{cr}}}{2} 
\end{equation}
indicating that the failure strain should be proportional to the particle spacing, $\lambda$, multiplied by 
$ \left[
\frac{I}{0.405 \pi l }
\right]^ \frac{1}{n+1}$.

\begin{figure}
\begin{center}
	\subfigure[Prismatic slip\label{fig-DuctilityPrismatic}]{\includegraphics[width=6cm]{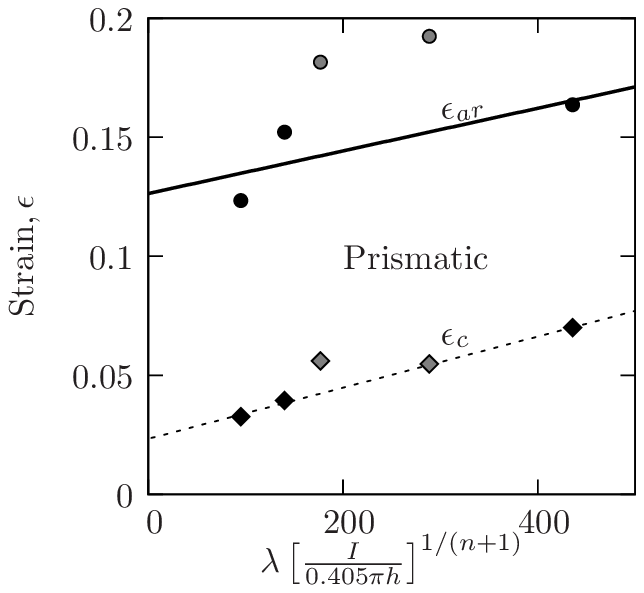}}
	\hfill
	\subfigure[Basal slip\label{fig-DuctilityBasal}]{\includegraphics[width=6cm]{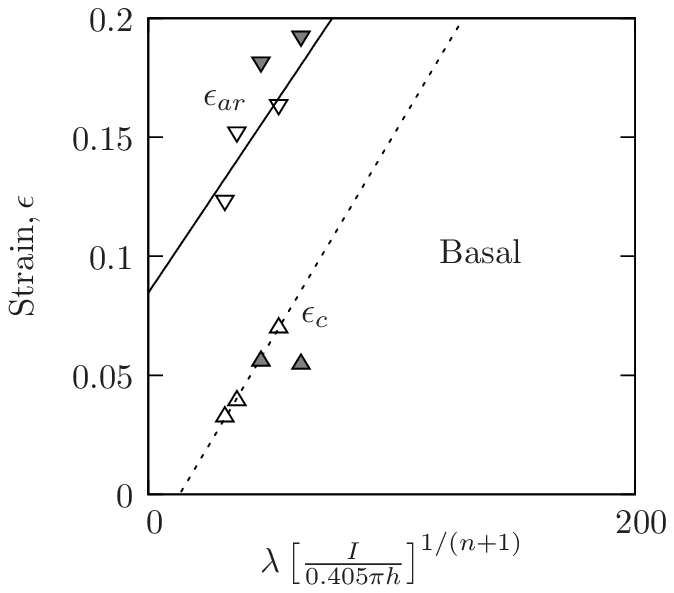}}
\caption{\label{fig-Ductility}
Tensile ductility of Mg-Zn and Mg-Zn-Y alloys versus \betap precipitate spacing on the (a) prismatic and (b) basal planes.
The true strain at  failure, $\epsilon_ar$ (diamonds) and onset of instability (i.e. uniform elongation, $\epsilon_c$) (circles) are plotted against the particle spacing multiplied by Ramberg-Osgood work hardening factor. 
Filled and open symbols represent values for Mg-Zn-Y and Mg-Zn, respectively.
}
\end{center}
\end{figure}

The relationship between failure strain and \betap interparticle spacing is shown in Figure~\ref{fig-Ductility}. 
The true strain to failure was calculated from the area reduction of fractured samples ($\epsilon_{ar}$) and plotted against the prismatic plane spacing  (multiplied by the work-hardening correction) in Figure~\ref{fig-DuctilityPrismatic}.
Figure~\ref{fig-DuctilityBasal} plots the same quantity against the basal plane spacing.
Filled symbols are for Mg-Zn-Y, while open symbols are for a Mg-3at.\%Zn alloy \cite{RosalieMgZn2012}.  
The maximum uniform strain (taken as the Consid\'ere strain, ($\epsilon_c$))  was plotted on the same axes.
This represents the strain required to initiate necking and provides an alternate measure for the tensile ductility of the samples.
Both $\epsilon_c$ and  $\epsilon_{ar}$ values followed a linear trend for both alloys, although there is substantial scatter in $\epsilon_{ar}$ values.

\subsection{The effectiveness of pre-ageing deformation}

Despite the large number of publications on Mg-Zn-Y alloys, to the best of the authors knowledge, none have examined the precipitation response as a function of the pre-ageing deformation. Effect of precipitation exclusively by slip is examined here, by using
extrusion to develop a fibre texture with the basal planes parallel to the extrusion axis \cite{AlSamman2010,Saxl1974}. 
This orientation precludes low-strain $\{10\overline{1}2\}$ twinning and ensures that pre-ageing deformation occurs by slip. 
This allowed the precipitate size and number density to be quantified as a function of the applied strain.

Without changing the composition of the alloy, the volume fraction of precipitates is increased by applying preageing deformation. With a simultaneous refinement of the size of the precipitates, a very substantial increase in the strength is achieved. This also highlights the effectiveness of precipitation over other strengthening contributions.

The yield strength improved substantially with 3\% pre-strain, but there was only a slight further improvement for 5\% pre-strain, suggesting that further increases in pre-ageing deformation may not be effective in increasing the mechanical properties. 
The relationship between precipitate spacing on the basal plane  and yield strength and ductility were similar to those observed in a Mg-3at.\%Zn alloy. 

\section{Conclusions}

The precipitation-strengthening response of a Mg-Zn-Y alloy were examined systematically for different levels of pre-ageing deformation. 
It was determined that;
\begin{itemize}
\item Without pre-ageing deformation the Mg-3at.\%Zn-0.5at.\%Y alloy shows a weak precipitate hardening response compared to binary Mg-Zn alloys, due to the formation of the ternary i-phase. 

\item Pre-ageing deformation of 3-5\% strain results in an increase in the volume fraction of \betap precipitates from approximately 0.5\% to  2.3\%  in the peak-aged condition.
This substantially increases the precipitation strengthening response of the alloy. 

\item Pre-ageing deformation reduces the average precipitate length from 475\,nm to 67\,nm and the diameter from 19.7\,nm 10.3\,nm. 
Precipitate lengths are within 11\% of those measured in Mg-Zn, while the
diameter values are 20-45\% greater, depending on pre-ageing deformation. 

\item The increase in \betap volume fraction and the refinement of the precipitation microstructure substantially increases the yield strength from  217\,MPa to 287\,MPa. 
The increase in yield strength is linear with reciprocal particle spacing on both  the basal and prismatic planes, as expected for Orowan looping of non-shearing particles.
\item Refinement of the precipitates substantially reduced the ductility. Elongation is reduced from 0.23 (0\% pre-strain) to 0.10 (5\% pre-strain).  
Uniform elongation and true strain to failure increase linearly with particle spacing and were equivalent to Mg-Zn alloys for identical particle spacing. 
\end{itemize}

\section*{Acknowledgements}

The authors  thank  Reiko Komatsu, Keiko Sugimoto and Toshiyuki Murao for assistance with sample preparation.


\begin{thebibliography}{10}
\expandafter\ifx\csname url\endcsname\relax
  \def\url#1{\texttt{#1}}\fi
\expandafter\ifx\csname urlprefix\endcsname\relax\def\urlprefix{URL }\fi
\expandafter\ifx\csname href\endcsname\relax
  \def\href#1#2{#2} \def\path#1{#1}\fi

\bibitem{KielbusElektron2007}
A.~Kielbus, Microstructure and mechanical properties of {E}lecktron 21 alloy
  after heat treatment, J Achiev Mater Manuf Eng 20~(1-2) (2007) 127--130.

\bibitem{WangCreep2001}
J.~Wang, L.~Hsiung, T.~Nieh, M.~Mabuchi, Creep of a heat treated {Mg-4Y-3RE}
  alloy, Materials Science and Engineering A 315~(1-2) (2001) 81--88.
\newblock \href {http://dx.doi.org/10.1016/S0921-5093(01)01209-6}
  {\path{doi:10.1016/S0921-5093(01)01209-6}}.

\bibitem{LuoReview2004}
A.~A. Luo, Recent magnesium alloy development for elevated temperature
  applications, International Materials Reviews {49}~({1}) ({2004}) {13--30}.
\newblock \href {http://dx.doi.org/10.1179/095066004225010497}
  {\path{doi:10.1179/095066004225010497}}.

\bibitem{BambergerReview2008}
M.~Bamberger, G.~Dehm, Trends in the development of new {Mg} alloys, Ann rev
  mater res {38} (2008) 505--533.
\newblock \href {http://dx.doi.org/10.1146/annurev.matsci.020408.133717}
  {\path{doi:10.1146/annurev.matsci.020408.133717}}.

\bibitem{Boehlert2007}
C.~J. Boehlert, The tensile and creep behavior of {Mg-Zn} alloys with and
  without {Y} and {Zr} as ternary elements, Journal of Materials Science 42
  (2007) 3674--3684.

\bibitem{BallPrangnell1994}
E.~A. Ball, P.~B. Prangnell, Tensile-compressive yield asymmetries in
  high-strength wrought magnesium alloys, Scripta Metallurgica et Materialia
  31~(2) (1994) 111--116.
\newblock \href {http://dx.doi.org/10.1016/0956-716X(94)90159-7}
  {\path{doi:10.1016/0956-716X(94)90159-7}}.

\bibitem{BohlenTexture2007}
J.~Bohlen, M.~R. Nuernberg, J.~W. Senn, D.~Letzig, S.~R. Agnew, The texture and
  anisotropy of magnesium-zinc-rare earth alloy sheets, Acta Mater. 55~(6)
  (2007) 2101--2112.
\newblock \href {http://dx.doi.org/10.1016/j.actamat.2006.11.013}
  {\path{doi:10.1016/j.actamat.2006.11.013}}.

\bibitem{StanfordTexture2008}
N.~Stanford, M.~Barnett, Effect of composition on the texture and deformation
  behaviour of wrought {Mg} alloys, Scr. Mat. 58~(3) (2008) 179--182.
\newblock \href {http://dx.doi.org/10.1016/j.scriptamat.2007.09.054}
  {\path{doi:10.1016/j.scriptamat.2007.09.054}}.

\bibitem{Li-Y-2011}
Y.~Li, F.~Xu, H.~Liu, Influence of {Y} element on microstructure and mechanical
  properties of {ZK60} alloy, Adv Mater Res 284-286 (2011) 1568--1573.
\newblock \href {http://dx.doi.org/10.4028/www.scientific.net/AMR.284-286.1568}
  {\path{doi:10.4028/www.scientific.net/AMR.284-286.1568}}.

\bibitem{RosalieMgZn2012}
J.~M. Rosalie, H.~Somekawa, A.~Singh, T.~Mukai, The effect of size and
  distribution of rod-shaped precipitates on the strength and ductility of a
  {Mg-Zn} alloy, Materials Science and Engineering A 539 (2012) 230--237.
\newblock \href {http://dx.doi.org/10.1016/j.msea.2012.01.087}
  {\path{doi:10.1016/j.msea.2012.01.087}}.

\bibitem{WeiPrec1995}
L.~Y. Wei, G.~L. Dunlop, H.~Westengen, Precipitation hardening of {Mg-Zn} and
  {Mg-Zn-RE} alloys, Metallurgical and Materials Transactions A 26A~(7) (1995)
  1705--1716.
\newblock \href {http://dx.doi.org/10.1007/BF02670757}
  {\path{doi:10.1007/BF02670757}}.

\bibitem{ShaoCalphad2006}
G.~Shao, V.~Varsani, Z.~Fan, Thermodynamic modelling of the {Y}-{Zn} and
  {Mg}-{Zn}-{Y} systems, CALPHAD 30~(3) (2006) 286--295.

\bibitem{HadornTexture2012}
J.~P. Hadorn, K.~Hantzsche, S.~Yi, J.~Bohlen, D.~Letzig, J.~A. Wollmershauser,
  S.~R. Agnew, Role of solute in the texture modification during hot
  deformation of {Mg}-rare earth alloys, Metallurgical and Materials
  Transactions A 43A~(4) (2012) 1347--1362.
\newblock \href {http://dx.doi.org/10.1007/s11661-011-0923-5}
  {\path{doi:10.1007/s11661-011-0923-5}}.

\bibitem{Singh2007}
A.~Singh, A.~P. Tsai, Structural characteristics of $\beta_1^\prime$
  precipitates in {Mg}-{Zn}-based alloys, Scr. Mat. 57 (2007) 941--944.
\newblock \href {http://dx.doi.org/10.1016/j.scriptamat.2007.07.028}
  {\path{doi:10.1016/j.scriptamat.2007.07.028}}.

\bibitem{Tsai1994}
A.~P. Tsai, A.~Niikura, A.~Inoue, T.~Masumoto, Y.~Nishida, K.~Tsuda, M.~Tanaka,
  Highly ordered structure of icosahedral quasicrystals in {Zn-Mg-RE} ({RE}=
  rare earth metals) systems, Philosophical Magazine Letters 70~(3) (1994)
  169--175.
\newblock \href {http://dx.doi.org/10.1080/09500839408240971}
  {\path{doi:10.1080/09500839408240971}}.

\bibitem{TsaiPhaseDiagram2000}
A.~P. Tsai, Y.~Murakami, A.~Niikura, The {Zn-Mg-Y} phase diagram involving
  quasicrystals, Philos. Mag. A 80~(5) (2000) 1043--1054.

\bibitem{Lepinoux2000}
J.~L\'{e}pinoux, Y.~Estrin, Mechanical behaviour of alloys containing
  heterogeneously distributed particles: Modelling with {D}elaunay
  triangulation, Acta Mater. 48~(17) (2000) 4337--4347.
\newblock \href {http://dx.doi.org/10.1016/S1359-6454(00)00181-6}
  {\path{doi:10.1016/S1359-6454(00)00181-6}}.

\bibitem{SinghTMS2008}
A.~Singh, Phase relations, formation and morphologies in {Mg-Zn-RE} ({RE=Y},
  rare earth) alloys, in: M.~O. Pekuleryuz, N.~R. Neelameggham, B.~R. S., E.~A.
  Nyberg (Eds.), Magnesium Technology, The Minerals, Metals and Materials
  Society, New Orleans, 2008, pp. 337--342.

\bibitem{Robson2011}
J.~D. Robson, N.~Stanford, M.~R. Barnett, Effect of precipitate shape on slip
  and twinning in magnesium alloys, Acta Mater. 59~(5) (2011) 1945--1956.
\newblock \href {http://dx.doi.org/10.1016/j.actamat.2010.11.060}
  {\path{doi:10.1016/j.actamat.2010.11.060}}.

\bibitem{AlSamman2010}
T.~Al-Samman, X.~Li, S.~G. Chowdhury,
Orientation  dependent slip and twinning during compression and tension of strongly textured magnesium {AZ31} alloy, Mater Sci Eng A 527~(15) (2010) 3450--3463,
\newblock \href {http://dx.doi.org/10.1016/j.msea.2010.02.008}
  {\path{doi:10.1016/j.msea.2010.02.008}}.

\bibitem{KoikeNonBasalSlip}
J.~Koike, T.~Kobayashi, T.~Mukai, H.~Watanabe, M.~Suzuki, K.~Maruyama,
  K.~Higashi, The activity of non-basal slip systems and dynamic recovery at
  room temperature in fine-grained {AZ31B} magnesium alloys, Acta Mater.
  51~(17) (2003) 2055--2065.

\bibitem{KoikeNonBasal2003a}
T.~Kobayashi, J.~Koike, T.~Mukai, M.~Suzuki, H.~Watanabe, K.~Maruyama,
  K.~Higashi, Anomalous activity of nonbasal dislocations in {AZ31} {Mg} alloys
  at room temperature, in: Y.~Kojima, T.~Aizawa, K.~Higashi, S.~Kamado (Eds.),
  Materials Science Forum, Vol. 419-4 of Materials science forum, 2003, pp.
  231--236.

\bibitem{Raeisinia2010}
B.~Raeisinia, S.~R. Agnew, Using polycrystal plasticity modeling to determine
  the effects of grain size and solid solution additions on individual
  deformation mechanisms in cast {Mg} alloys, Scr. Mat. 63~(7) (2010) 731--736.
\newblock \href {http://dx.doi.org/10.1016/j.scriptamat.2010.03.054}
  {\path{doi:10.1016/j.scriptamat.2010.03.054}}.

\bibitem{Blankenship1993}
C.~P. Blankenship, E.~Hornbogen, E.~A. Starke, Predicting slip behavior in
  alloys containing shearable and strong particles, Materials Science and
  Engineering A 169~(1-2) (1993) 33--41.

\bibitem{FriedelDislocations1964}
J.~Friedel, Dislocations, Pergamon, London, 1964.

\bibitem{shercliff:1990a}
H.~R. Shercliff, M.~F. Ashby, A process model for age hardening of aluminium
  alloys-{I.} the model, Acta Metallurgica et Materialia 38~(10) (1990)
  1789--1802.

\bibitem{StanfordBarnett2009}
N.~Stanford, M.~R. Barnett, Effect of partictles on the formation of
  deformation twins in a magnesium-based alloy, Materials Science and
  Engineering A 516 (2009) 226--234.

\bibitem{WuLaves2010}
M.-M. Wu, L.~Wen, B.-Y. Tang, L.-M. Peng, W.-J. Ding, First principles study of
  elastic and electronic properties of {MgZn$_2$} and {ScZn$_2$2} phase in
  {Mg-Sc-Zn} alloy, J. Alloys. Comp. 506 (2010) 412--417.
\newblock \href {http://dx.doi.org/10.1016/j.jallcom.2010.07.018}
  {\path{doi:10.1016/j.jallcom.2010.07.018}}.

\bibitem{Lepinoux2001}
J.~L{\'{e}}pinoux, Y.~Estrin, Mechanical behaviour of alloys containing stable
  or evolving heterogeneous arrangements of particles, Journal de Physique IV
  11~(PR4) (2001) 357--364.

\bibitem{AvedesianMg99}
M.~M. Avedesian, H.~Baker (Eds.), Magnesium and magnesium alloys, {ASM}
  speciality handbook, {ASM}, 1999, p.~79.

\bibitem{chan1995}
K.~Chan,A  fracture model for hydride-induced embrittlement, Acta Metallurgica et  Materialia 43~(12) (1995) 4325--4335.
\newblock \href {http://dx.doi.org/10.1016/0956-7151(95)00133-G}
  {\path{doi:10.1016/0956-7151(95)00133-G}}.


\bibitem{LiuZhang2004}
G.~Liu, G.~Zhang, X.~Ding, J.~Sun, K.~Chen, The influences of multiscale-sized
  second-phase particles on ductility of aged aluminum alloys, Metallurgical
  and Materials Transactions A 35 (2004) 1725--1734.
\newblock \href {http://dx.doi.org/10.1007/s11661-004-0081-0}
  {\path{doi:10.1007/s11661-004-0081-0}}.

\bibitem{Ashby1970}
M.~F. Ashby, The deformation of plastically inhomogeneous materials,
  Philosophical Magazine 21~(170) (1970) 399--.
\newblock \href {http://dx.doi.org/10.1080/14786437008238426}
  {\path{doi:10.1080/14786437008238426}}.

\bibitem{RambergOsgood1943}
W.~Ramberg, W.~R. Osgood, Description of stress-strain curves and their
  parameters, Tech. Rep. 902, NACA (1943).

\bibitem{Hutchinson1968}
J.~W. Hutchinson, Singular  behaviour at the end of a tensile crack in a hardening material, J Mech Phys Solid 16~(1) (1968) 13--31.
\newblock \href {http://dx.doi.org/10.1016/0022-5096(68)90014-8}
  {\path{doi:10.1016/0022-5096(68)90014-8}}.

\bibitem{LiuSun2005}
G.~Liu, J.~Sun, C.-W. Nan, K.-H. Chen, Experiment and multiscale modeling of
  the coupled influence of constituents and precipitates on the ductile
  fracture of heat-treatable aluminum alloys, Acta Mater. 53~(12) (2005)
  3459--3468,
\newblock \href {http://dx.doi.org/10.1016/j.actamat.2005.04.002}
  {\path{doi:10.1016/j.actamat.2005.04.002}}.

\bibitem{Dowling1987}
N.~E. Dowling,
J-integral  estimates for cracks in infinite bodies, Eng Fracture Mech 26~(3) (1987)
  333--348.
\newblock \href {http://dx.doi.org/10.1016/0013-7944(87)90016-6}
  {\path{doi:10.1016/0013-7944(87)90016-6}}.

\bibitem{Saxl1974}
I.~Saxl, I.~Haslinerova, Double prismatic slip in textured {Mg}-2 percent {Be}
  alloy, Czech J Phys {B 24}~({12}) ({1974}) 1351--1361.
\newblock \href {http://dx.doi.org/10.1007/BF01589812}
  {\path{doi:10.1007/BF01589812}}.

\end{thebibliography}
\end{document}